\documentclass[aps,prb,twocolumn,superscriptaddress,showpacs,floatfix,longbibliography]{revtex4}
\usepackage{amssymb}
\usepackage{graphicx}
\usepackage{amsmath}
\usepackage{times}
\usepackage{color}
\usepackage{setspace}
\usepackage{bm}
\usepackage{epsfig}
\usepackage{amsmath}
\usepackage[breaklinks=true,colorlinks,citecolor=blue,linkcolor=blue,urlcolor=blue]{hyperref}
\def\be{\begin{equation}}
\def\ee{\end{equation}}

\def\ber{\begin{eqnarray}}
\def\eer{\end{eqnarray}}

\begin{document}
\newcommand {\bqa} {\begin{eqnarray}}
\newcommand {\eqa} {\end{eqnarray}}
\newcommand {\ba} {\ensuremath{b^\dagger}}
\newcommand {\no} {\nonumber}
\newcommand{\kk}{{\bf k}}

\title{Anisotropic transport of normal metal-barrier-normal metal junctions in monolayer phosphorene}

\author{Sangita De Sarkar}
\email{tpsds@iacs.res.in}
\affiliation{Theoretical Physics
Department, Indian Association for the Cultivation of Science,
Jadavpur, Kolkata 700032, India.}

\author{Amit Agarwal}
\email{amitag@iitk.ac.in}
\affiliation{Department of Physics, Indian Institute of Technology Kanpur, Kanpur 208016, India}

\author{K. Sengupta}
\email{tpks@iacs.res.in}
\affiliation{Theoretical Physics
Department, Indian Association for the Cultivation of Science,
Jadavpur, Kolkata 700032, India.}

\begin{abstract}

We study transport properties of a phosphorene monolayer in the
presence of single and multiple potential barriers of height $U_0$
and width $d$, using both continuum and  microscopic lattice models,
and show that the nature of electron transport along its armchair
edge ($x$ direction) is qualitatively different from its counterpart
in both conventional two-dimensional electron gas with
Schr\"odinger-like quasiparticles and graphene or surfaces of
topological insulators hosting massless Dirac quasiparticles.  We
show that the transport, mediated by massive Dirac electrons, allows
one to achieve collimated quasiparticle motion along $x$ and thus
makes monolayer phosphorene an ideal experimental platform for
studying Klein paradox. We study the dependence of the tunneling
conductance $G \equiv G_{xx}$ as a function of $d$ and $U_0$, and
demonstrate that for a given applied voltage $V$ its behavior
changes from oscillatory to decaying function of $d$ for a range of
$U_0$ with finite non-zero upper and lower bounds, and provide
analytical expression for these bounds within which $G$ decays with
$d$. We contrast such behavior of $G$ with that of massless Dirac
electrons in graphene and also with that along the zigzag edge ($y$
direction) in phosphorene where the quasiparticles obey an effective
Schr\"odinger equation at low energy. We also study transport
through multiple barriers along $x$ and demonstrate that these
properties hold for transport through multiple barrier as well.
Finally, we suggest concrete experiments which may verify our
theoretical predictions.

\end{abstract}
%
%
\maketitle
\section{Introduction}

Two dimensional crystals  composed of single or few atomic layers
are the focus of intense research currently, on account of their
remarkable optical, electronic and mechanical properties. Moreover
such materials, as shown in recent years \cite{revgr1,revti1}, serve
as test bed for Dirac physics. This property of these materials
arises from the fact that their low-energy quasiparticles obey an
effective Dirac-like equation. Such quasiparticles lead to a host of
unconventional thermodynamic and transport properties. Examples of
such unconventional properties seen in the context of graphene and
topological insulators include unconventional quantum Hall effect
\cite{hallref1}, unusual Kondo effect \cite{kondoref1}, and
unconventional transport properties \cite{transref1,transref2}.
Indeed, the latter property serves as one of the key aspects of
Dirac (relativistic) materials which distinguishes them from
conventional materials whose quasiparticles obey Schr\"odinger
equation. For example, the former class of materials display
oscillatory behavior of the transmission through a potential barrier
as a function of the barrier height or width; this behavior is in
complete contrast to an exponentially decaying transmission function
found for the latter class
\cite{History_klein,Klein_nature,Robinson}.

Phosphorene \cite{ph-1, ph-review} is another such Dirac material
which is being actively investigated for electronic and other
applications \cite{ph-tr1, ph-tr2, ph-tr3} on account of its
anisotropic electronic \cite{ph-bs1-dft}, thermal \cite{ph-thermal1,
ph-thermal2} and optical properties \cite{ph-bs1-dft}. A hallmark of
monolayer phosphorene is its anisotropic band-structure, which
displays an almost flat parabolic Schr\"odinger like dispersion
along the zigzag edge ($\Gamma-Y$ direction) and a predominantly
Dirac like dispersion along the armchair edge ($\Gamma-X$
direction)\cite{ph-bs1-dft, ph-bs2-tb, ph-bs2-tb-2,Doh, ph-bs3-kp,
ph-bs3-tb, ph-bs4-tb}. Such anisotropic bandstructure distinguishes
phosphorene from other Dirac materials such as graphene or
topological insulator surface where the effective Dirac theory is
massless and isotropic. Accordingly, the transport properties across
single or multiple potential barriers in phosphorene are expected to
be different depending on the orientation of the barrier relative to
the $X-Y$ plane. The anisotropy of the low energy dispersion in
phosphorene thus offers an opportunity to simultaneously probe the
relativistic as well as the non-relativistic nature of electrons in
a transport experiment based on a normal-barrier-normal (NBN)
junctions. Depending on the orientation of the phosphorene monolayer
with respect to the measuring electrodes, it is expected to display
an oscillatory behavior in the transmission function as a function
of the barrier strength as well as the barrier height along the
armchair direction, and an exponential decaying transmission
function along the zigzag direction.  However, a detailed
theoretical investigation of such properties has not been yet
carried out.

It is the aim of this paper to highlight these anisotropic transport
signatures of phosphorene across a NBN junction with different
orientations. More specifically, we study the transport properties
of quasiparticles in monolayer phosphorene in the presence of single
and multiple potential barriers oriented along $\Gamma-X$
(subsequently referred to as the $x$ direction). The main results
that we obtain from such a study are as follows. First, we show that
the transport shows several signatures consistent with a gapped
Dirac quasiparticle; this is in sharp contrast to the case when the
barrier is along $y$ ($\Gamma-Y$ direction) for which the transport
properties conform to those due to Schr\"ondinger quasiparticle.
Second, we show that for a barrier along the $x$ direction, the
dominant contribution to the transport comes from the quasiparticles
which impinge on the barrier at near-normal incidence provided the
applied voltage is close to the bottom of the conduction band,
leading to collimated transport of electrons. The degree of such
collimation can be tuned by the external applied voltage $V$. Third,
we find that this property, which is experimentally inaccessible in
gapless Dirac materials such as graphene and topological insulators,
allows us to tune to a regime where the conductance $G_{xx} \simeq
G$ mimics the behavior of normal transmission amplitude. This leads
to the possibility of observing signature of Klein paradox through
measurement of $G$. Fourth, we find that both the normal
transmission $T(k_y=0)$ and $G$ displays oscillatory or decaying
behavior as a function of the barrier width $d$ depending on the
relative strength of the dimensionless barrier potential $\zeta =
U_0/2m$ and the applied voltage $\eta=eV/2m$, where $2m$ is the mass
gap; the behavior of $T(k_y=0)$ and $G$ changes from oscillatory to
a monotonically decreasing function of $d$ for $\zeta_1 \le \zeta
\le \zeta_2$. We also show analytically that $\zeta_1=\eta$ and
$\zeta_2=\eta+1$ for $k_y=0$ using a continuum approximation to the
lattice model of phosphorene and demonstrate that the phenomenon
described above persists beyond the continuum approximation used to
obtain the expression of $\zeta_1$ and $\zeta_2$. We point out that
such transport behavior is qualitatively different from their
counterparts in both conventional Schr\"odinger and gapless Dirac
materials. Fifth, we study transport of phosphorene quasiparticles
through multiple barriers (each of height $U_0$ and width $d$) along
$x$ and demonstrate that all of the above-mentioned features derived
for transport through a single barrier holds in the multiple barrier
case. The main difference between the two manifests itself in
additional peaks in normal transmission $T$ or conductance $G$ as a
function of $d$ in the oscillatory regime ($ \zeta_1\ge \zeta\ge
\zeta_2$) for the multiple barrier case; we provide an analytical
explanation for this phenomenon for $n=2$ barriers. Finally, we
discuss experiments which can test our theory.

The manuscript is organized as follows. In Sec. \ref{sec1}, we
outline the band structure of phosphorene and chart out the
continuum Hamiltonian used for transport calculation. Such
calculations for the single potential barrier along $x$ is discussed
in Sec.\ \ref{sec2}, while transport through multiple barriers along
$x$ is discussed in Sec.\ \ref{sec3}. Finally, we discuss our
results, chart out possible experiments which can test them, and
conclude in Sec.\ \ref{sec4}.

\section{Low energy effective Hamiltonian of Phosphorene}
\label{sec1}

\begin{figure}[t]
\begin{center}
\includegraphics[width=1.0 \linewidth]{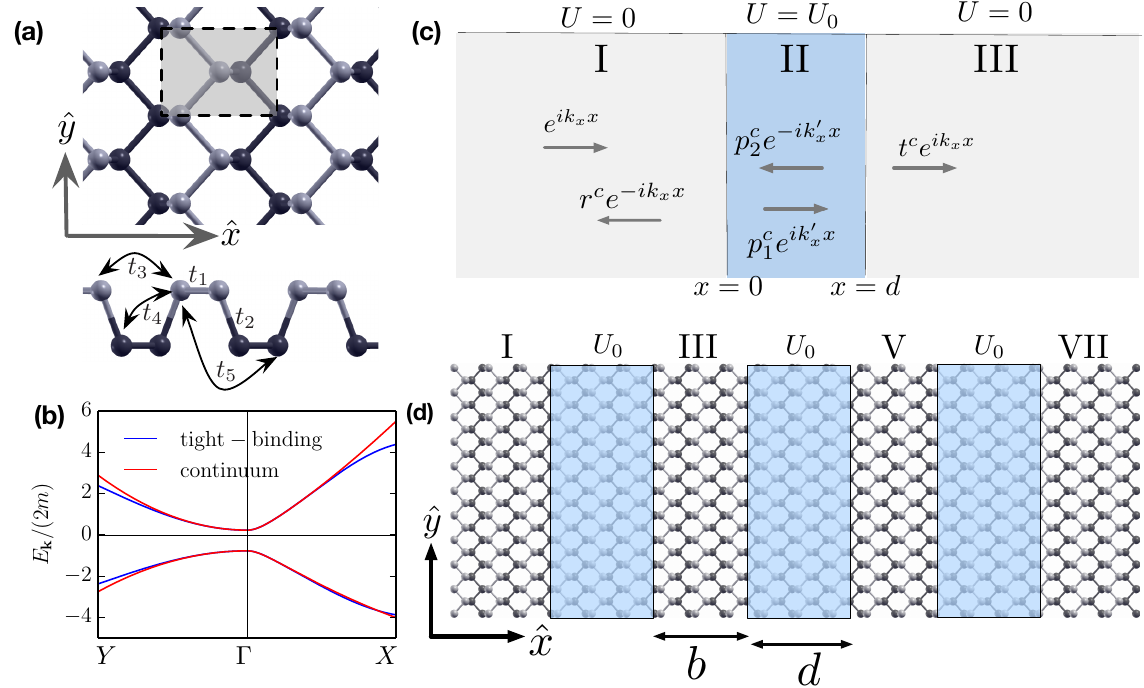}
\caption{ a) The top and side-view (along the armchair edge) of the
lattice structure of monolayer phosphorene, with the shaded area
depicting the unit cell, and the side view highlighting the hopping
parameters used for calculating the tight-binding Hamiltonian. b)
The low energy anisotropic bandstructure of monolayer phosphorene,
using both tight-binding and the continuum approximations. (c)
Schematic of scattering across a rectangular barrier along $x$ in
phosphorene, for a wave incident from the left. (d) Multiple (three)
rectangular barriers in phosphorene, with applied potential $U_0$,
arranged periodically along its armchair direction (along ${x}$).
Note that the barrier regions are labeled by even numbers.
\label{fig1}}
\end{center}
\end{figure}

The band-structure of monolayer phosphorene, is well known from
ab-initio calculations \cite{ph-bs1-dft, ph-bs2-tb}, and it has been
used to construct effective low energy Hamiltonian using several
approaches such as  ${\bf k}\cdot {\bf p}$ method \cite{ph-bs3-kp},
the tight-binding approach \cite{ph-bs2-tb, ph-bs2-tb-2, ph-bs3-tb,
ph-bs4-tb,Doh}, and the methods of invariants \cite{ph-bs-inv1}. All
of these methods yield qualitatively similar band structure. Thus
for this paper we use the two band Hamiltonian which is obtained
from the four band tight-binding Hamiltonian on a discrete lattice,
making use of the $D_{2h}$ symmetry \cite{ph-bs3-tb} and expanding
around the $\Gamma$ point. The origin of such $D_{2h}$ symmetry can
be understood in the following manner.  The unit cell of
phosphorene, shown in Fig.\ \ref{fig1}(a), contains four phosphorus
atoms, such that the upper and the lower layers each contain two of
these atoms. This leads to the $D_{2h}$ point group invariance which
constitutes invariance under a shift along the plane of the layer
combined with exchange of the layer indices. It is therefore
sufficient to consider a reduced unit cell of two atoms either in
the upper or in the lower layer since the transfer energy of atoms
in these layers are identical.

Our starting point is the effective two-band tight-binding model
based on Refs.~[\onlinecite{ph-bs2-tb,ph-bs3-tb}]. This
tight-binding model is given by \bqa H=\sum_{<i,j>}t_{ij}
c_i^{\dagger} c_j~, \label{tbh0} \eqa where the summation runs over
the lattice sites, $t_{ij}$ is the hopping matrix element between
$i^{\rm th}$ and $j^{\rm th}$ sites, and $c_j$ is the annihilation
operator of an electrons at site $j$. The corresponding two-band
Hamiltonian can be written in momentum space as \cite{ph-bs3-tb}
\begin{eqnarray}
H_1 &=& \sum_{\kk} \psi_{\kk} H_{1 \kk} \psi_{\kk}~,~
H_{1 \kk} = f_{\kk} I + g_{1\kk} \tau_x + g_{2\kk}~
\tau_y \label{hamk1}~, \nonumber \\
f_{\kk} &=& 4t_4 \cos[\sqrt{3}k_x/2] \cos[k_y/2]~, \\
g_{1 \kk} &=& 2 t_1 \cos[k_x/(2 \sqrt{3})] \cos[k_y/2] + t_2
\cos[k_x/\sqrt{3}]
\nonumber\\
&& + 2t_3 \cos[5k_x/(2 \sqrt{3})] \cos[k_y/2] +
t_5\cos[2k_x/\sqrt{3}]~,
\nonumber \\
g_{2 \kk} &=& - 2 t_1 \sin[k_x/(2 \sqrt{3})] \cos[k_y/2] + t_2
\sin[k_x/\sqrt{3}]
\nonumber\\
&& + 2t_3 \sin[5k_x/(2 \sqrt{3})] \cos[k_y/2] -
t_5\sin[2k_x/\sqrt{3}]~, \nonumber
\end{eqnarray}
where $t_i$ for $i= 1...5$ are the different transfer matrix
elements indicated in Fig.\ \ref{fig1}(a) with  $t_1=-1.220$ eV,
$t_2=3.665$ eV, $t_3=-0.205$ eV, $t_4=-0.105$ eV and $t_5=-0.055$
eV, $\tau_{x,y}$ are the Pauli matrices in the band basis,
$\psi_{\bf k} = (c_{1\bf k}, c_{2 \bf k})$ is the two component fermion
field with $c_{1(2) \bf k}$ being the annihilation operators
corresponding to electrons of the two phosphorene atoms from the unit
cell, $I$ denotes the identity matrix, and we have
scaled wavevectors $\kk= (k_{x},k_{y})$ by the corresponding lattice
lengths $a_{x}$ and $a_y$.
The energy spectrum of this Hamiltonian is given by
\begin{eqnarray}
E_{\kk \pm} &=& f_{\kk} \pm \sqrt{g_{1\kk}^2 + g_{2 \kk}^2}~,
\label{eigenen1}
\end{eqnarray}
where the $+(-)$ sign in the subscript indicates conduction
(valence) band.  A plot of these bands as a function of $k_x$ (for
$k_y=0$) and $k_y$ (for $k_x=0$) is shown in Fig.\ \ref{fig1}(b).
The corresponding eigenvectors are given by
\begin{eqnarray}
\psi_{\kk \pm} &=& (1, \lambda_0 e^{i \theta_{\kk}})^{T}/\sqrt{2}~,
\quad \tan \theta_{\kk} = g_{2\kk}/g_{1\kk}~, \label{eigenfn1}
\end{eqnarray}
where $\lambda_0 = {\rm Sgn}(E-f_{\kk})$ and ${\rm Sgn}$ denotes
the signum function.

The Hamiltonian $H_{1 \kk}$ may be simplified in the low-energy
low-momentum or continuum limit. Using the identities $\cos x \to 1
-x^2/2$ and $ \sin x \to x$, one can obtain the continuum version of
Eq.~\eqref{hamk1}: $H_{1\kk} \to  H_{1 \kk}^c$,  to be
\begin{eqnarray}
H^c_{1\kk} &=& f^c_{\kk} + \left(m +\alpha k_x^2 + \beta
k_y^2\right)\tau_x + \gamma k_x \tau_y~, \label{tbh1}
 \end{eqnarray}
where $\mu$ is the chemical potential, and $f^c_{\kk}=t_4\left(4-3
k_x^2/2- k_y^2/2\right)-\mu$, with $t_4=-0.105$eV. The other
parameters in Eq.~\eqref{tbh1}, can be expressed in terms of $t_i$
in a straightforward manner and are given\cite{ph-bs3-tb} by  $m = 2
(t_1 + t_3) + t_2 + t_5= 0.76$ eV, $\gamma = (-t_1 + t_2 + 5 t_3 - 2
t_5)/\sqrt{3}= 2.29$eV, $\alpha = -(t_1+2 t_2 + 25t_3 + 8t_5)/12=
-0.045$eV, and  $\beta = -(t_1 + t_3)/4= 0.36$eV. Here we note that
$\gamma \gg \alpha$ which allows one to have an approximately linear
dispersion in $k_x$ for small $\kk$. The eigenvalues and
eigenfunctions of $H_{1\kk}^c$ are given by
\begin{eqnarray}
E^{c}_{\kk \pm} &=& f^c_{\kk} \pm  \Big[ \left(m +\alpha k_x^2
+ \beta k_y^2\right)^2 + \gamma^2 k_x^2 \Big]^{1/2}, \label{endisp1} \\
\psi_{\pm}^{c} (\kk) &=& \frac{1}{\sqrt{2}} (1,   \lambda_0^c e^{
i\phi_{\kk}})^{T}, \,\,  \tan\phi_{\kk} = \frac{\gamma k_x}{m
+\alpha k_x^2 + \beta k_y^2}~, \nonumber
\end{eqnarray}
and  $\lambda_0^c = {\rm Sgn}(E-f^c_{\kk})$. From Fig.\
\ref{fig1}(b), we find that the energy bands of the continuum model
match with those of the lattice model. We note that in the continuum
limit, since $\gamma \gg \alpha$, the Hamiltonian is linear (Dirac
like) in $k_x$ but parabolic (Schr\"odinger like) in $k_y$;
consequently, the nature of the quasiparticle transport and the
effect of a potential barrier depends crucially on the its
orientation in the $x-y$ plane. We shall discuss this phenomenon in
detail in the next section.

\section{ Transport across a single barrier: NBN junction}
\label{sec2}

In this section, we present an analysis of transport across a single
barrier. We first present the general formalism for both the lattice
model (Eq.\ \eqref{hamk1}) and its continuum approximation (Eq.\
\eqref{tbh1}) in Sec.\ \ref{form1}. The numerical results of the analysis of the formalism
developed in Sec.\ \ref{form1} is presented in Sec.\ \ref{result1}.

\subsection{Formalism}
\label{form1}

In this section, we study ballistic transport in phosphorene
monolayer in the presence of a single barrier of strength $U_0$ and
width $d$ as shown in Fig.~\ref{fig1}(c). We will primarily be
focussed on transport in the $x$ direction, which is likely to have
signatures of Dirac quasiparticles.

Let us first discuss the formalism for calculating transmission and
conductance along $x$ using the two band lattice Hamiltonian of
Eq.~\eqref{hamk1}. The wave function for the particles with
transverse momenta $k_y$ and energy $E= \mu + eV$ (which we choose
to lie in the conduction band), where $\mu$ is the chemical
potential and $V$ is the applied bias voltage, moving along $\pm x$
direction can be obtained by diagonalization of Eq.~\eqref{hamk1}.
The corresponding eigenfunctions can be read off from
Eq.~\eqref{eigenfn1}; for an electron moving along $\pm x$ (right
(+) and left (-) moving electrons) with momenta $k_y$, it is given
by $\psi^{\pm}(x) e^{i k_y y}$ with \ber
\psi^{\pm}(x) &=& \frac{1}{\sqrt{2}} \left(\begin{array}{c} 1 \\
\lambda_0 e^{\pm i\theta_{\kk}}\end{array}\right)~ e^{\pm i k_{1x}
x}~, \label{wav0} \eer where $\kk = (k_{1x}, k_y)$ and $k_{1x}
\equiv k_{1x}(E;k_y)$ is obtained from the solution of
\begin{eqnarray}
E &=& eV + \mu  = f_{\kk} + \sqrt{g_{1\kk}^2 + g_{2 \kk}^2}
\label{keq2at}~,
\end{eqnarray}
We first note that the $\pm$ sign here in the superscript indicates
right(left) moving electrons in contrast to those in the subscript
in Sec.\ \ref{sec1} which indicated conduction and valence band
energies. Since for transport we shall focus only on electrons in
the conduction band in region I, we omit the subscript indicating
conduction or valence band for brevity. We also note that
Eq.~\eqref{keq2at} admits two solution for $k_{1x}$ corresponding to
any given energy $E$ and transverse momentum $k_y$. One of these
solutions is real and represents a propagating wave while the other
is imaginary and represents an evanescent wave. While the
expectation of the current operator and hence the conductance $G$
receives contribution from the propagating wave, the formalism that
we use requires that the evanescent wave solutions are treated at an
equal footing. In what follows, we denote the solutions, obtained by
numerical solution of Eq.~\eqref{keq2at}, as $\pm k_{1x}^R \equiv
\pm k_{1x}^R(E;k_y)$ and $\pm i k_{1x}^{I} \equiv \pm i k_{1x}^I(E,
k_y)$ with $k_{1x}^I >0$. In terms of these wavevectors, one can
write the wavefunction in region I (which extends in the region
$x<0$ as seen in Fig.~\ref{fig1}(c)) as
\begin{eqnarray}
\psi_1(x) &=& \frac{1}{\sqrt 2} \Bigg[ e^{i k_{1x}^R x}\left(\begin{array}{c} 1 \\
\lambda_0 e^{i\theta_{1}} \end{array}\right) + r_1 e^{-i k_{1x}^R x}\left(\begin{array}{c} 1 \\
\lambda_0 e^{i\theta_{2}} \end{array}\right) \nonumber\\
&& + r_2 e^{k_{1x}^I x}\left(\begin{array}{c} 1 \\
\lambda_0 e^{i\theta_{3}}\end{array}\right) \Bigg] ~,\label{wav1}
\end{eqnarray}
where $\theta_1 \equiv \theta_{(k_{1x}^R, k_y)}$, $\theta_2 =
\theta_{(-k_{1x}^R, k_y)}$, and $\theta_3 = \theta_{(-i k_{1x}^I, k_y)}$
is obtained using Eq.~\eqref{eigenfn1}. In Eq.~\eqref{wav1} we have
kept only the decaying evanescent modes, and $r_1$ and $r_2$ denotes
reflection coefficients corresponding to the propagating and the
evanescent modes respectively.

The wavefunction in region II (as shown in Fig.~\ref{fig1}(c))
consists of left and right propagating modes as well as evanescent
waves which decays and grows within $0 \le x\le d$. Here depending
on the magnitude of the applied barrier potential $U_0$ and $k_y$,
the longitudinal wavevector may be obtained from either the valence
or the conduction band energy expressions (Eq.\ \ref{eigenen1}).
This is in contrast to the situation in regions I and III where they
are necessarily obtained using energy expressions for the conduction
band. The longitudinal wavevector for these modes are obtained from
the solution of
\begin{eqnarray}
E &=& eV + \mu  = U_0 + f_{\kk} \pm \sqrt{g_{1\kk}^2 + g_{2
\kk}^2} \label{keq3at}
\end{eqnarray}
and we denote these wavevectors as $ \pm k_{2x}^R$ and $ \pm i
k_{2x}^I$. In terms of these, one can write the wavefunction in
region II as
\begin{eqnarray}
\psi_2(x) &=& \frac{1}{\sqrt 2} \Bigg[ p_1 e^{i k_{2x}^R x}\left(\begin{array}{c} 1 \\
\lambda'_0 e^{i\theta'_{1}} \end{array}\right) + p_2 e^{-i k_{2x}^R x}\left(\begin{array}{c} 1 \\
\lambda'_0 e^{i\theta'_{2}} \end{array}\right) \nonumber\\
&& + q_1 e^{k_{2x}^I x}\left(\begin{array}{c} 1 \\
\lambda'_0 e^{i\theta'_{3}}\end{array}\right) + q_2 e^{-k_{2x}^I x}\left(\begin{array}{c} 1 \\
\lambda'_0 e^{i\theta'_{4}}\end{array}\right) \Bigg], \nonumber \\  \label{wav2}
\end{eqnarray}
where $\theta'_1 = \theta_{(k_{2x}^R,k_y)}$, $\theta'_2 = \theta_{(-k_{2x}^R,k_y)}$,
$\theta'_3 = \theta_{(-ik_{2x}^I,k_y)}$ $\theta'_4 = \theta_{(ik_{2x}^I,k_y)}$, 
and $\lambda'_0 = {\rm sgn}(E-U_0-f_{\kk})$. In Eq.~\eqref{wav2},
$p_1$ and $p_2$ are coefficients corresponding to the left and the
right propagating mode, and $q_1$ and $q_2$ are the coefficients
corresponding to decaying and growing evanescent mode. Note that
since region II extends within a finite span of $0 \le x \le d$, the
solution with the growing evanescent mode is admissible in this
region.

Finally, we consider the wavefunction in region III which extends
for $x>d$. In this region, one finds
\begin{eqnarray}
\psi_3(x) &=& \frac{1}{\sqrt 2} \Bigg[ t_1 e^{i k_{1x}^R x}\left(\begin{array}{c} 1 \\
\lambda_0 e^{i\theta_{1}} \end{array}\right) + t_2 e^{-k_{1x}^I x}\left(\begin{array}{c} 1 \\
\lambda_0 e^{i\theta_{4}}\end{array}\right) \Bigg], \nonumber\\
\label{wav3}
\end{eqnarray}
where $\theta_4 = \theta_{(ik_x^I,k_y)}$, $t_1$ and $t_2$ are the
transmission amplitude corresponding to the propagating and the
evanescent mode. Note that in this region, similar to region I, only
the decaying evanescent mode is admissible.

To obtain the reflection and transmission coefficients, we use the
standard procedure of imposing the continuity condition for the wavefunction as well as the current
at $x=0$ and $x=d$. This leads to the conditions
\begin{eqnarray}
\psi_1(0) &=& \psi_2(0), \quad \psi_2(d)= \psi_3(d) \nonumber\\
\hat v_{x} \psi_1(0) &=& \hat v_{x} \psi_2(0), \quad \hat v_{x}
\psi_2(d) = \hat v_{x} \psi_3(d) \label{bc1}
\end{eqnarray}
where the velocity operator can be obtained using Eq.~\eqref{hamk1} and is given by
\begin{eqnarray}
\hat v_{x} &\equiv& \partial H_{\kk}/\partial k_x \nonumber \\ &=& \tau_x \partial_{k_x} g_{1 \kk} + \tau_y \partial_{k_x} g_{2
\kk} + \partial_{k_x} f_{\kk} I \label{velop}~,
\end{eqnarray}
with the substitution $k_x \to - i \partial_x$. Substituting Eqs.~ \eqref{wav1}-\eqref{wav3} in Eq.~\eqref{bc1},
we can then obtain the following eight equations,
\begin{widetext}
\begin{eqnarray}
&& 1+r_1+r_2-p_1-p_2-q_1-q_2=0~, \nonumber\\
&& \lambda_0 \left( e^{i\theta_1}+ r_1 e^{i\theta_2} + r_2
e^{i\theta_3} \right) - \lambda^\prime_0 \left( p_1
e^{i\theta^\prime_1}+ p_2 e^{i\theta^\prime_2} + q_1
e^{i\theta^\prime_3}+ q_2 e^{i\theta^\prime_4} \right)=0~,
\nonumber\\
&&p_1 e^{i k^R_{2x}d}+ p_2 e^{-i k^R_{2x}d}+ q_1 e^{ k^I_{2x}d}+ q_2
e^{- k^I_{2x}d}- t_1 e^{i k^R_{1x}d}- t_2 e^{-k^I_{1x}d}=0~,
\nonumber\\
&& \lambda^\prime_0 \left( p_1e^{i(\theta^\prime_1 + k^R_{2x}d)} +
p_2 e^{i(\theta^\prime_2-k^R_{2x}d)} + q_1
e^{i\theta^\prime_3+k^I_{2x}d }+ q_2 e^{i\theta^\prime_4-k^I_{2x}d}
\right) - \lambda_0 \left( t_1 e^{i(\theta_1 +k^R_{1x})d} + t_2
e^{i\theta_4 -k^I_{1x}d}\right)=0~, \nonumber\\
&&A(k^R_{1x}) + r_1 A(-k^R_{1x}) + r_2 A(-i k^I_{1x}) -p_1
A(k^R_{2x}) -p_2 A(-k^R_{2x}) - q_1 A(-i k^I_{2x})- q_2A(i
k^I_{2x})=0~, \nonumber\\
&& B(k^R_{1x}) + r_1 B(-k^R_{1x}) + r_2 B(-i k^I_{1x}) -p_1
B(k^R_{2x}) -p_2 B(-k^R_{2x}) - q_1 B(-i k^I_{2x})- q_2 B(i
k^I_{2x})=0~, \label{latbc1} \\
&& p_1 A(k^R_{2x}) e^{i k^R_{2x}d}
+p_2 A(-k^R_{2x}) e^{-i k^R_{2x}d}+ q_1 A(-i k^I_{2x}) e^{k^I_{2x}d}
+q_2A(i k^I_{2x})e^{-k^I_{2x}d}-t_1A(k^R_{1x}) e^{i k^R_{1x}d}-t_2
A(ik^I_{1x}) e^{-k^I_{1x}d} =0~, \nonumber\\
&& p_1 B(k^R_{2x}) e^{i k^R_{2x}d}+p_2 B(-k^R_{2x}) e^{-i
k^R_{2x}d}+q_1 B(-i k^I_{2x})e^{k^I_{2x}d} + q_2 B(i
k^I_{2x})e^{-k^I_{2x}d}-t_1 B(k^R_{1x}) e^{i k^R_{1x}d}-t_2 B(i
k^I_{1x}) e^{-k^I_{1x}d}=0~, \nonumber
\end{eqnarray}
\end{widetext}
where $g_{\kk} = g_{1 \kk} - i g_{2\kk}$ and the quantities
$A$ and $B$ are defined as
\begin{eqnarray}
A(k_x) &=& \partial_{k_x} f_{\kk}+ \lambda_{k_x}
e^{i\theta_{k_x}} \partial_{k_x} g_{\kk}~, \nonumber\\
B(k_x)&=& \partial_{k_x} f_{\kk}+ \lambda_{k_x} e^{i\theta_{k_x}}
\partial_{k_x} g^{\ast}_{\kk}~. \label{ABdef}
\end{eqnarray}
Here $\theta_{k_x}\equiv \theta_{(k_x,k_y)}$ can assume values
$\theta_1 .. \theta_4$ depending on values of $k_x$: $k_x= \pm
k_{1(2)x}^R$ or $\pm i k_{1(2)x}^I$, $\lambda_{k_x}$ takes values of
$\lambda_0$ or $\lambda'_0$ depending on $k_x = k_{1 x}$ or
$k_{2x}$, and we have suppressed the $k_y$ and energy dependence of
$A$ and $B$ for clarity. Note that the presence of the evanescent
waves (complex solutions for $k_x$) is essential for unique solution
of Eq.~\eqref{latbc1}; without them, we would have an overdetermined
set of equations.

Equation~\eqref{latbc1} is solved numerically to obtain the reflection
and the transmission coefficients $r_1$ and $t_1$. The transmission
probability $T(eV;k_y)$ can be computed in terms of these as
$T(eV;k_y) = |t_1(k_y)|^2$. Note that $t_2$ which corresponds to the
transmission amplitude of the evanescent mode decays exponentially
in region III away from the barrier and hence it does not contribute to
the transmission for $ L \gg d$, where $L$ is the system size and
the lead is placed at $x=L$. To compute the conductance in the
presence of a single barrier we first note that the current operator
along $x$ is given by $ J_x =(i e/\hbar) \psi^{\ast} {\hat v}_x \psi$
and thus the current flowing in regions I and III are given by
\begin{eqnarray}
J_x^{\rm in} &=& (i e/\hbar) \psi_1^{\ast} {\hat v}_x \psi_1~, ~~~{\rm and}~~~ \nonumber \\
J_x^{\rm tr} &=& (i e/\hbar) \psi_{3}^{\ast} {\hat v}_x \psi_{3}~.
\label{curre1}
\end{eqnarray}
The conductance $G$ of the system can then be computed as
\begin{eqnarray}
G(eV) &=& G_0 \int_{-k_y^{\rm max}}^{k_y^{\rm max}} \frac{dk_y}{2
\pi} \frac{J_x^{\rm tr}}{J_x^{\rm in}} ~,\label{condlatxx}
\end{eqnarray}
where $G_0=e^2 L_y/\hbar$, and $k_y^{\rm max}$ denotes the maximum
transverse momenta for which Eq.~\eqref{endisp1} admits a real
solution for $k_x(E,k_y)$ and the limits of integration are decided
by the fact that the dispersion in Eq.~\eqref{eigenen1} is symmetric
in $k_y$.

Having described the formalism to calculate the transmission and
conductance for the two band lattice Hamiltonian, we now focus our
attention on the low energy and small momentum limit, for which the
system is described by $H^c_{\kk}$. In what follows we shall neglect
the $\alpha k_x^2$ term in the expression of $H^c_{\kk}$; this is
justified by the fact that for low momenta $m, \gamma k_x \gg \alpha
k_x^2$ and $k_x^2 \le 1$ so thats the $k_x$ dependence of
$f^c_{\kk}$ may be neglected. The motivation for solving the problem
within this approximation is two-fold. First, as we shall see, the
conductance and transmission coefficients computed within this
approximation matches those from the exact lattice model at low
applied voltages and second, this approximation yields an analytic
expression for the transmission coefficient which allow further
insight into the transport properties of the system.

For electrons described by Eq.~\eqref{tbh1} with $\alpha=0$, let us
consider a wave incident on the barrier with energy $E$ and
transverse wavevector $k_y$. The wavefunction of the electron can be
computed using Eqs.~\eqref{tbh1}-\eqref{endisp1} as $\psi_{\rm in}^c
(x;k_y) = \exp(i k_x x) (1, \lambda_0^c \exp[i \phi_1])/\sqrt{2}$,
where  $\lambda_0^c = {\rm Sgn}(E- t_4(4 - k_y^2/2))$,
$\tan[\phi_1]= \gamma k_x/(m+\beta k_y^2)$ and $k_x$ is given by
\begin{eqnarray}
k_x = \gamma^{-1} \sqrt{(E-t_4(4 -k_y^2/2))^2 -(m+\beta k_y^2)^2}~.
\label{kxcont}
\end{eqnarray}
The reflected wavefunction in this case is given by
$\psi_r^c(x;k_y)= \exp(-i k_x x) (1,\lambda_0^c \exp[-i
\phi_1])/\sqrt{2}$.  Thus the wave function in region I (see Fig.\
\ref{fig1}(c)) can be written as
\begin{eqnarray}
\psi^c_1 (x;k_y) &=& \psi_{\rm in}^c + r^c \psi_r^c \label{wavcont1}~.
\end{eqnarray}
We note that the continuum approximation does not support the
evanescent modes found in the lattice formulation. This feature is
consistent with the fact that for the effective Dirac-like
Hamiltonian with linear dispersion that we found within this
approximation, we only need continuity of the wavefunctions at the
barrier edges ($x=0$ and $x=d$); the continuity of the derivative of
the wavefunction is no longer necessary since we are dealing with
linear differential operators along $x$. This situation is in
contrast to Schr\"odinger electrons for which both the wavefunction
and its derivative needs to be continuous across the barrier
\cite{transref1,transref2}.

In the barrier region, the wavefunction can be expressed as
\begin{eqnarray}
\psi_2^c(x;k_y) &=& \frac{1}{\sqrt 2} \Bigg[ p^c_1 e^{i k'_{x} x}\left(\begin{array}{c} 1 \\
\lambda^{'c}_0 e^{i\phi'_1} \end{array}\right) \nonumber\\
&& + p^c_2 e^{-i k'_{x} x}\left(\begin{array}{c} 1 \\
\lambda^{'c}_0 e^{-i\phi'_1} \end{array}\right)
\Bigg]~,\label{wavcoont2}
\end{eqnarray}
where $\lambda^{'c}_0 = {\rm Sgn}[E- U_0- t_4(4 -k_y^2/2)]$,
$\tan[\phi'_1] = \gamma k'_x/(m+\beta k_y^2)$ and $k'_x$ is given by
\begin{eqnarray}
k'_x = \gamma^{-1} \sqrt{[E-U_0-t_4(4 -k_y^2/2)]^2 -(m+\beta
k_y^2)^2}. \label{wavecont2}
\end{eqnarray}
Finally, in region III, the transmitted electron wavefunction is
given by
\begin{eqnarray}
\psi_3^c(x;k_y) &=& \frac{t^c}{\sqrt{2}} e^{i k_{x} x}\left(\begin{array}{c} 1 \\
\lambda^c_0 e^{i\phi_1} \end{array}\right).
\end{eqnarray}
The reflection and the transmission coefficients can be determined
using standard procedure of demanding wavefunction continuity at the
interfaces at $x=0$ and $x=d$: $\psi^c_1|_{x=0}= \psi^c_{2}|_{x=0}$
and $\psi^c_{2}|_{x=d}= \psi^c_{3}|_{x=d}$. This yields
\begin{eqnarray}
&& 1+r^c-p^c_1-p_2^c =0~, \nonumber\\
&& \lambda^c_0 (e^{i\phi_1} + r^c e^{-i \phi_1}) - \lambda^{'c}_0
(p_1^c e^{i\phi'_1} + p_2^c  e^{-i \phi'_1}) =0~, \nonumber\\
&&p_1^c e^{i k'_x d} + p_2^c e^{-i k'_x d} - t^c e^{i k_x x} =0~,
\\
&& \lambda^{'c}_0 p_1^c e^{i (k'_x d +\phi')} + \lambda^{'c}_0 p_2^c e^{-i (k'_x d +\phi')}- \lambda^c_0 t^c
e^{i (k_x d +\phi)} =0~. \nonumber \label{bccont1}
\end{eqnarray}
A solution of Eq.~\eqref{bccont1} leads to the transmission and
reflection amplitudes within the continuum approximation,
\begin{eqnarray}
r^c&=&\frac{|B_2|^2+2i\lambda^c_0 B_1\sin(\phi_1)+B_1 B_2 e^{2ik'_{x}d}}{|B_1|^2 e^{2ik'_{x}d}-|B_2|^2}\no\\
t^c&=&\frac{-4\lambda^c_0\lambda^{'c}_0 \sin(\phi_1) \sin(\phi'_1)
e^{i(k'_{x}-k_x)d}}{|B_1|^2 e^{2i k'_{x}d}-|B_2|^2} \label{rtcont1}
\end{eqnarray}
where $B_1=\lambda^{'c}_0 e^{i\phi'_1}-\lambda^c_0 e^{i\phi_1}$ and
$B_2=\lambda^c_0 e^{i\phi_1}-\lambda^{'c}_0 e^{-i\phi'_1}$. The
transmission probability $T \equiv T(E;k_y) = |t^c|^2$  is then given
by \bqa
T&=&\frac{4\sin^2(\phi_1)\sin^2(\phi'_1)}{4\sin^2(\phi_1)\sin^2(\phi'_1)
+|B_1|^2|B_2|^2\sin^2{k'_x d}} \label{transcont} ~.\eqa The
conductance $G$ may then be obtained using
\begin{eqnarray}
G(V) = G_0 \int \frac{dk_y}{2\pi} T(eV; k_y) \label{contcond1}~.
\end{eqnarray}
Note that Eq.~\eqref{contcond1} is a simplified version of
Eq.~\eqref{condlatxx}, for the case when the incoming and the
outgoing regions  across the barrier are  identical. We shall
analyze the results obtained from the formalism developed in this
section for both the lattice model and its continuum approximation
in  Sec.\ \ref{result1}.

Before ending this section, we analyze Eq.~\eqref{transcont} in the
thin barrier limit where $U_0, m \gg eV$ and $d \to 0$ with a fixed
ratio $\zeta = U_0/2m$, $\eta = eV/2m \ll 1$ and $\chi_0  =U_0 d/\gamma$.
In this limit, Eq.~\eqref{transcont} may be further simplified to obtain
\begin{eqnarray}
T_1 &=& \frac{1}{1+\alpha
\sin^2\chi}, \quad {\rm where} \quad \alpha = {\mathcal N}/{\mathcal D}~, \quad \nonumber\\
 \chi &=& \chi_0 \left[(\eta-\zeta)(\eta+1-\zeta)/\zeta^2
\right.
\nonumber\\
&& \left.+ k^2_y(t_4-2\beta+2 t_4(\eta-\zeta))/(4m\zeta^2)
\right]^{1/2}~, \nonumber \\
{\mathcal N} &=& (1 + \beta k_y^2/m) \left[ |2(\zeta-\eta) -1 -t_4
k_y^2/2m| \right. \nonumber\\
&& \left. - \lambda_0 \lambda'_0 |2\eta + 1 +t_4 k_y^2/2m|
\right]^2~,\nonumber\\
{\mathcal D} &=& \frac{16 \zeta^2 \chi^2}{\chi_0^2} \left [\eta + k_y^2(t_4/2 - \beta
+\eta t_4)/(2m)\right]~.
\label{tblt1}
\end{eqnarray}
\begin{figure}[t]
\begin{center}
\includegraphics[width=0.48 \linewidth]{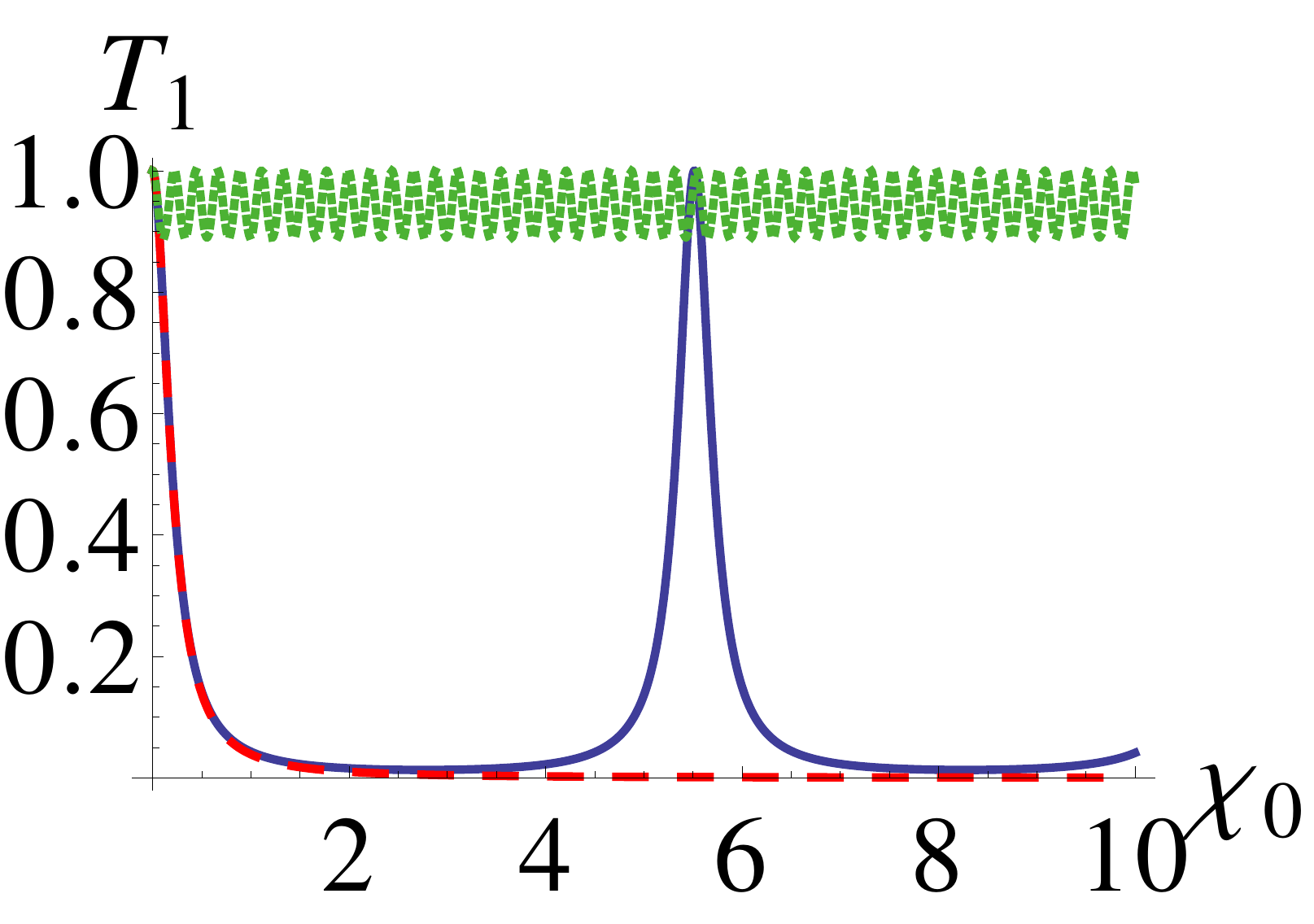}
\includegraphics[width=0.48 \linewidth]{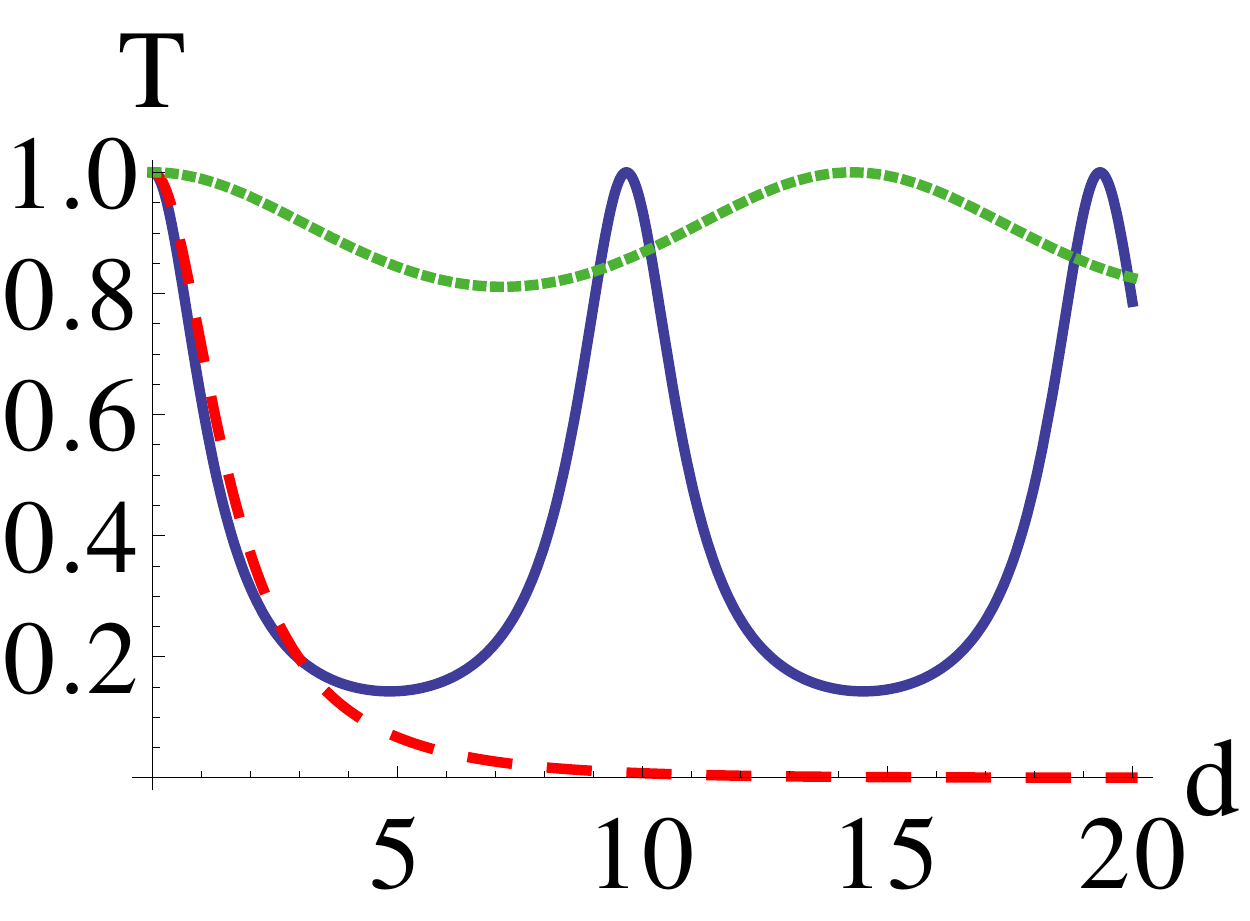}
\caption{ Plot of the normal transmission $T_1(k_y=0)$ in the thin
barrier limit as a function of the effective barrier strength
$\chi_0 = Ud/\gamma$, for several representative values of $\zeta$
for $\eta=0.01$ (left panel). In the left panel, $\zeta=1.5$ (blue
solid line), $1$ (red dashed line) and $0.005$ (green dotted line).
In the right panel, $T(k_y=0)$ (Eq.\ \ref{transcont}) is plotted as
a function of $d$ with $\eta = 0.3$ for $\zeta=1.5$ (blue solid
line), $1.2$ (red dashed line) and $0.2$ (green dotted line). The
plots show distinct oscillatory behavior for $\zeta >\zeta_1$ and
$\zeta< \zeta_2$, which turns into a decaying behavior for
$\zeta_2<\zeta <\zeta_1$ in both the panels. This is the regime
where the wavevector in the barrier region becomes imaginary. In
both the panels we have chosen $\mu = 4 t_4 + m$  so that $\eta = 0$
corresponds to the bottom of the conduction band. See text for
details. \label{fig2}}
\end{center}
\end{figure}

Equation~\eqref{tblt1} can be analyzed to obtain several characteristics
of the transmission coefficient $T_1$. We first note that as $\eta ,
\zeta \to \infty$ (which means $m \to 0$) and $k_y=0$, $T_1 \to 1$
which reproduces Klein paradox result for massless Dirac fermions
seen in graphene. Second for $k_y=0$, $\chi \to \chi_1 =
\chi_0\sqrt{(\eta-\zeta)(\eta+1-\zeta)}/\zeta$, and consequently
$T_1$ as a function of the effective barrier strength $\chi_0$
oscillates with a frequency $f_0 =
\sqrt{(\eta-\zeta)(\eta+1-\zeta)}/(\pi \zeta)$ as $U_0$ or $d$ is
varied; thus knowing $U_0$ and $d$, one can estimate the mass of
single-layer phosphorene by measuring the frequency of such
oscillation. Third, for $k_y=0$ as $U_0$ is tuned such that
$\zeta_1< \zeta < \zeta_2$, where
\begin{eqnarray}
\zeta_1 = \eta, \quad \zeta_2= \eta+1 , \label{critzeta}
\end{eqnarray}
$\chi$ becomes imaginary and hence $T_1$ changes from being
oscillatory to a decaying function of $d$. For $\zeta=\zeta_{1/2}$,
we find that
\begin{eqnarray}
T_1(k_y=0) &=& \left(1+  \frac{m U_0^2}{ 2 eV \gamma^2} d^2\right)^{-1}. \label{tcrit}
\end{eqnarray}
Such a qualitative change in $T_1$, which is generally not seen in
other massless Dirac materials, may serve as an accurate measurement
of the mass gap in phosphorene. It is to be noted that while $\zeta
\le \zeta_1=\eta$ indicates that the applied voltage is larger than
the barrier height which is expected to lead to non-decaying
behavior of $T$ (and hence $G$) with $d$, the oscillatory behavior
of $T$ for $\zeta>\zeta_2$ is a property of Dirac nature of the
phosphorene electrons. The difference of such transport from that in
graphene follows from the decaying behavior of $T$ for $\zeta_1 \le
\zeta\le \zeta_2$; such a behavior is absent for transport in
graphene and topological insulators where $T$ and $G$ are
oscillatory functions of $d$ for any $\zeta > \eta$. We find that
this property is not a consequence of the thin barrier limit can
also be seen from Eqs.~\eqref{transcont} and \eqref{wavecont2}.
Clearly the decaying behavior arises only when the wavevector in the
potential region, is imaginary. Fourth, we note that the
contribution to the transmission modes comes from transverse momenta
modes for which $k_x^2 >0$ away from the barrier. This requires the
condition $ k_y \le k_y^c \equiv \sqrt{eV/(\beta - t_4/2)}$, where
we have set the chemical potential so that $V=0$ corresponds to the
bottom of the conduction band. Thus by tuning the applied voltage
one may reach a regime where the conductance
\begin{eqnarray}
G_c = G_0 \int \frac{dk_y}{2 \pi} T_1
\end{eqnarray}
receives its contribution only from the quasi particles with near
normal incidence. This leads to collimated transport; furthermore
since the $k_y$ dependence of $\chi$ in this regime is negligible,
the oscillation frequency of $G$ with the barrier width $\chi$ would
mimic that of $T_1$. This enables us to realize a setup where Klein
paradox could be realized via measurement of $G$; we note that it is
impossible to reach this regime in gapless Dirac systems such as
graphene. We shall study the feasibility of this proposition in
details in the next section. Finally, we note that in contrast to
graphene, the presence of a finite mass gap allows us to tune
transmission through the barrier. This can be seen by inspecting
Eq.~\eqref{wavecont2}; we find that there are no propagating modes
inside the barrier, for $k_y \ge k_y^{'c} =
\sqrt{(U_0-eV-2m)/(\beta+t_4/2)}$. The key point is that the value
of $k_y^{'c}$ can tuned to zero by choosing $U_0 \simeq 2m +eV$
which allows one to tune, particularly for large $d$, transmission
through the barrier by tuning $U_0$.

\subsection{Results}
\label{result1}

In this subsection, we shall chart out the results corresponding to
the theory of transport of monolayer phosphorene electrons through a
single barrier along the $x$ direction developed in Sec.\ \ref{form1}.
We shall first analyze the results for the continuum model (Eq.~\eqref{hamk1})
and then compare its prediction with those obtained from the lattice
model (Eq.~\eqref{tbh1}).

To this end, we first plot the normal transmission $T_1(k_y=0)$ as a
function of the barrier width $d$ for several representative values
of $\zeta$ (barrier height) for fixed $\eta$ (incoming energy) as
shown in Fig.\ \ref{fig2}. We note that for $\zeta> \zeta_2=1+\eta$
(or alternately $U_0 > eV + 2m$), $T_1$ displays oscillatory
behavior with increasing $d$ which is similar to that of massless
Dirac-like electrons as seen in graphene \cite{Klein_nature} and in
contrast to that of conventional Schr\"odinger electrons with
parabolic dispersion for which the transmission probability
decreases monotonically with $d$. However, in contrast to graphene,
$T_1$ shows a decaying behavior as a function of $d$ for $\zeta_1
\le \zeta \le \zeta_2$ [or alternately for $ eV \le U_0 \le eV +
2m$; (Eq.~\eqref{critzeta})]. As $\zeta$ is further decreased below
$\zeta_1=\eta$ (or $U_0 < eV$), $T_1$ becomes an oscillatory
function of $d$ with frequency given by $\pi/k'_x$ (green dashed
plot in the left panel of Fig.\ \ref{fig2}). A similar behavior
is seen for $T$ (Eq.\ \ref{transcont}) as shown in the right panel
of Fig.\ \ref{fig2}. We note that this leads to tunability of
transmission which is a result of both the band structure of
phosphorene and the presence of the mass gap $m$; such behavior is
therefore absent in graphene.

\begin{figure}[t]
\begin{center}
\includegraphics[width=0.48 \linewidth]{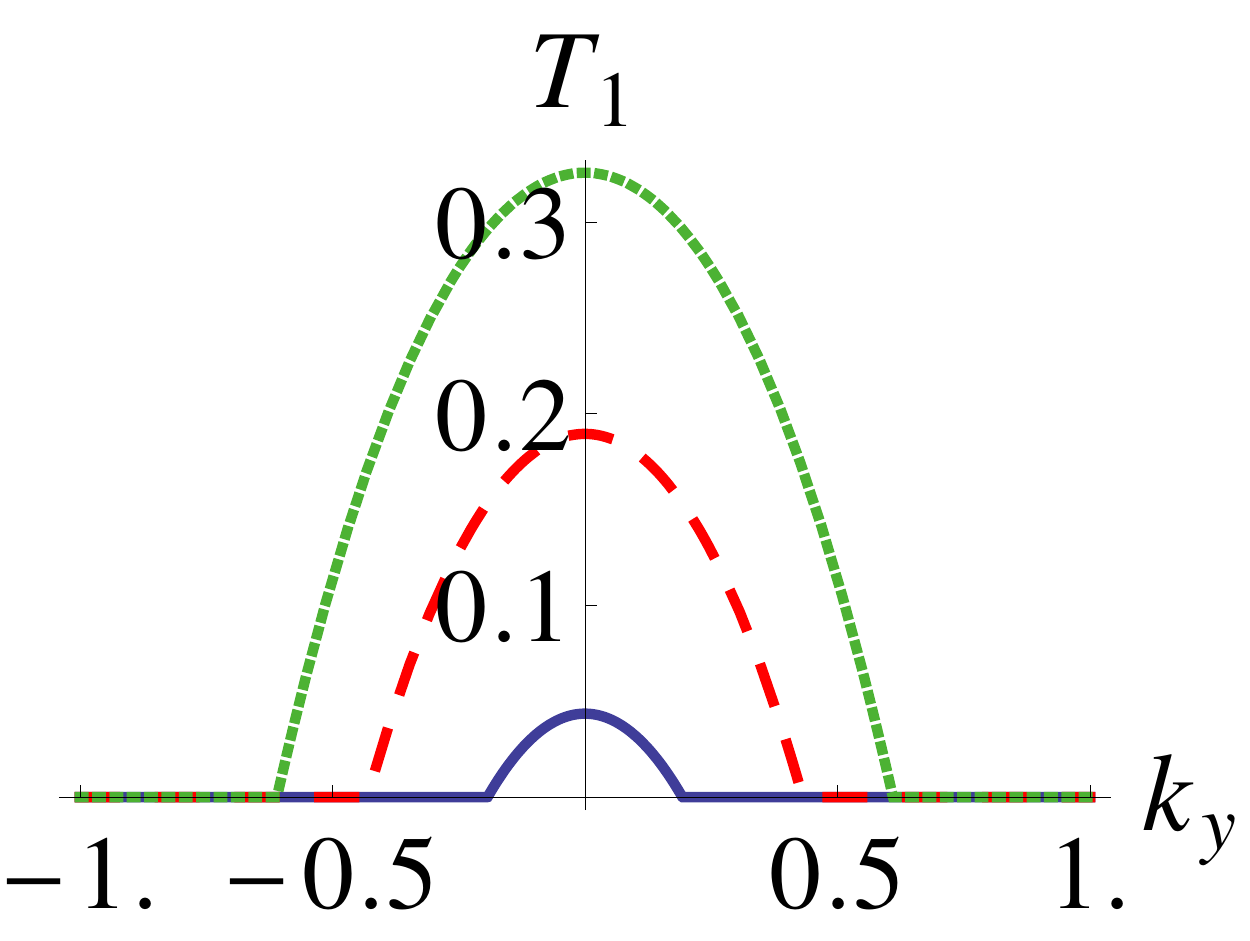}
\includegraphics[width=0.48 \linewidth]{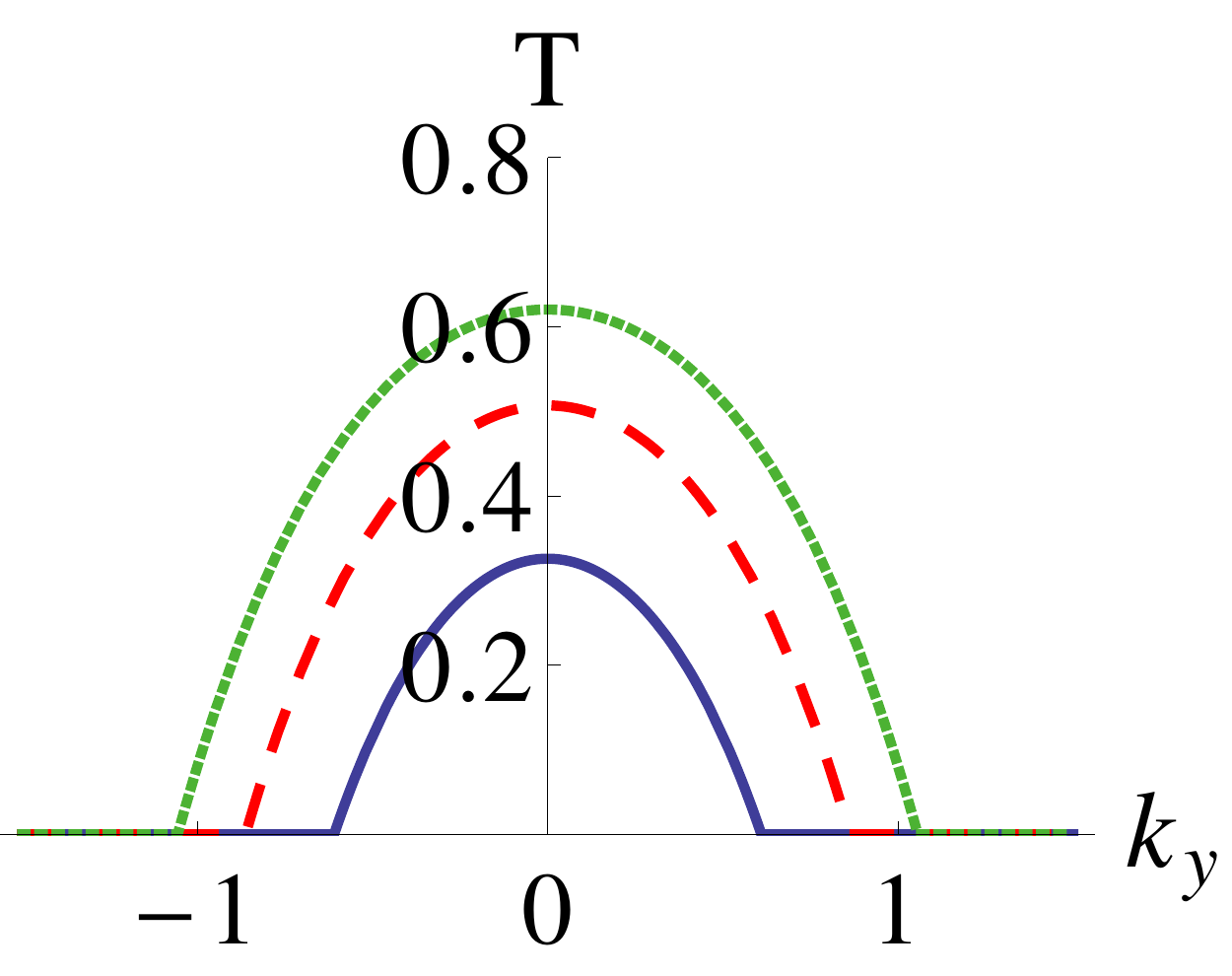}
\caption{Plot of the transmission probability as a function of $k_y$
for several values of $\eta$. Here $\zeta=1.5$, $d=1$ (in the unit
of lattice spacing). In the left panel, $T_1(k_y)$ (thin barrier
limit) is plotted as a function of $k_y$ for $\eta=0.01$ (blue solid
line), $0.05$ (red dashed line) and $0.1$ (green dotted line). In
the right panel, $T(k_y)$ (Eq.\ \ref{transcont}) is plotted as a
function of $k_y$ for $\eta=0.1$ (blue solid line), $0.2$ (red
dashed line) and $0.3$ (green dotted line). All other parameters
have the same value as in Fig.\ \ref{fig2}. The plot shows
increasingly collimated transport with decreasing $\eta$. See text
for details. \label{fig3}}
\end{center}
\end{figure}

\begin{figure}[t]
\begin{center}
\includegraphics[width=0.48 \linewidth]{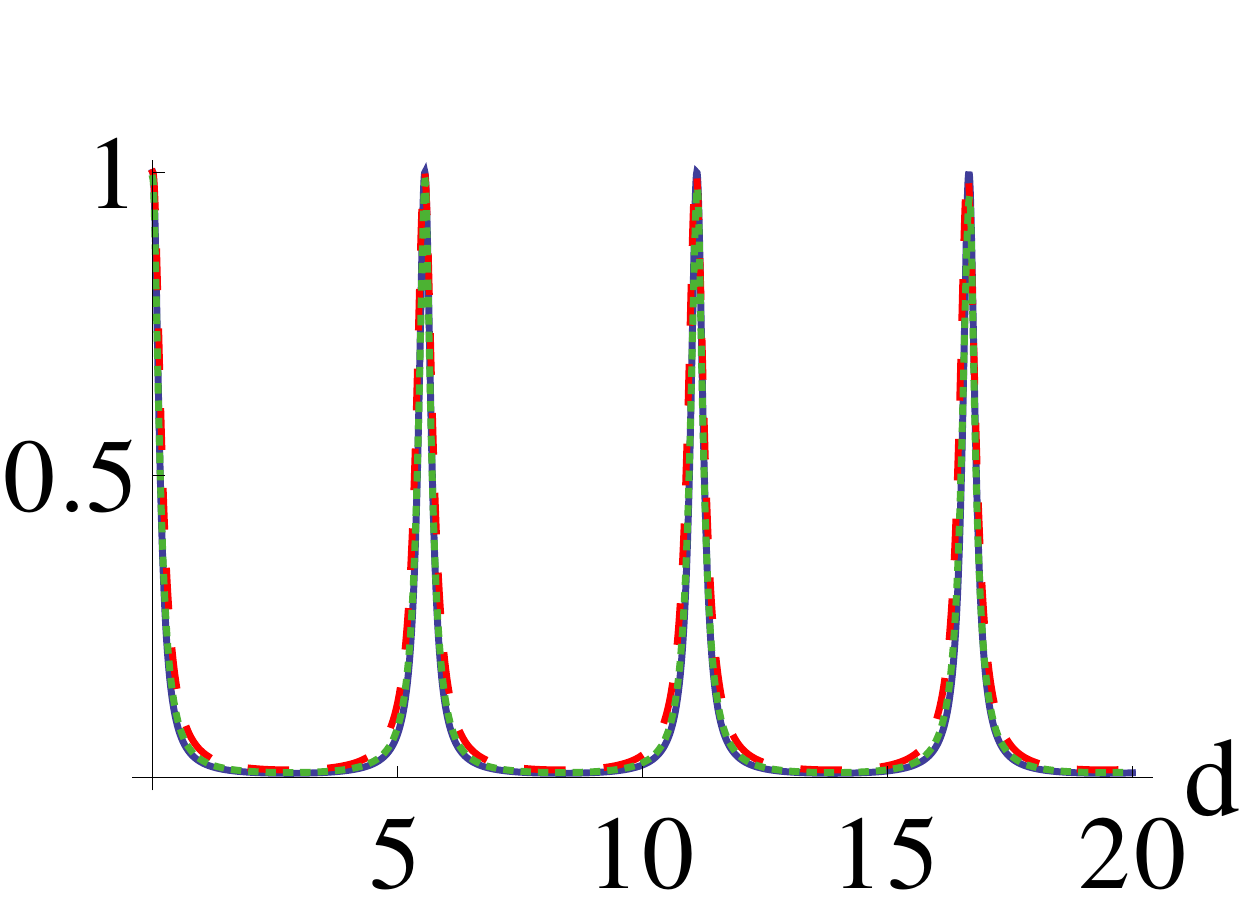}
\includegraphics[width=0.48\linewidth]{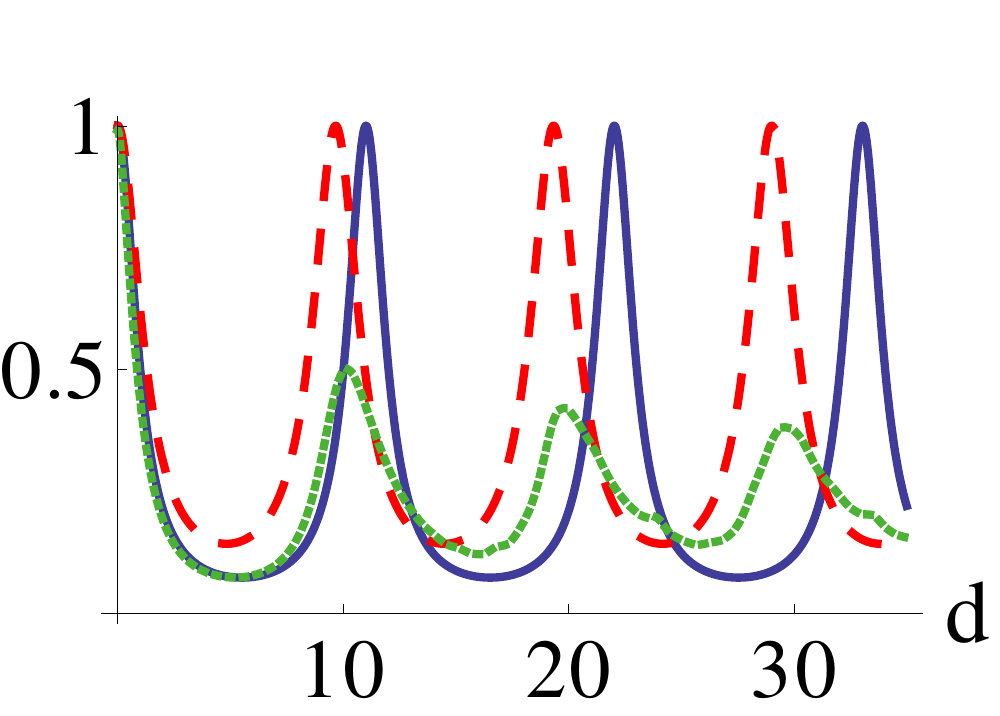}
\caption{ Plot of the conductance $G/G_0$ and transmission
probabilities as a function of $d$ for $\eta=0.01$ in the thin
barrier limit (left panel) and $\eta=0.3$ (right panel). In the left
panel, $T_1(k_y=0)$ (blue solid line), $T_1(k_y=\pi/25)$ (red dashed
line) and $G/G_0$ (green dotted line). In the right panel,
$T(k_y=0)$ (blue solid line), $T(k_y=\pi/6)$ (red dashed line) and
$G/G_0$ (green dotted line). For both plots, all other parameters
are same as in Fig.\ \ref{fig3}. We note that for small $\eta$ (near
the band bottom), $G/G_0$ mimics the normal transmission; however,
the behavior of these quantities are quite different for larger
$\eta$. \label{fig4}}
\end{center}
\end{figure}

Next, we plot $T_1(k_y)$ as a function of $k_y$ for a fixed $d=1$,
$\mu= 4t_4 +m$, and $\zeta=1.5$ for several representative values of
$\eta$ as shown in the left panel of Fig.\ \ref{fig3}. As we
discussed in the last section, the transport becomes increasingly
collimated as the applied voltage $V$ is tuned towards the bottom of
the conduction band (which in our notation corresponds to $\eta=0$)
since lesser number of $k_y$ modes satisfies Eq.~\eqref{kxcont} for
real values of $k_x$. This feature, which holds beyond thin
barrier limit as shown from the plot of $T(k_y)$ (Eq.\
\ref{transcont}) in the right panel of Fig.\ \ref{fig3}, can be
shown to be related to the relative flatness of phosphorene bands
near the band bottom along the $y$-direction in contrast to that in
the $x$-direction; this naturally leads to collimated behavior.
Thus, at sufficiently low $\eta$, $T_1(k_y=0)$ dominates the
conductance $G$ as shown in the left panel of Fig. \ref{fig4}. At
low $\eta$, the oscillatory behavior of $T_1$ as a function of $d$,
which is a signature of Klein paradox for gapped Dirac systems
\cite{Klein_nature}, may therefore be observed via measurement of
tunneling conductance $G$. We note that such a measurement would be
impossible in graphene since one needs to be very close to the Dirac
point to observe this phenomenon where the density of state is
extremely small. Such an equivalence between $G$ and $T(k_y=0)$
(Eq.\ \ref{transcont}) is lost for larger $\eta$ as shown in the
right panel of Fig.\ \ref{fig4}.

\begin{figure}[t]
\begin{center}
\includegraphics[width=0.48 \linewidth]{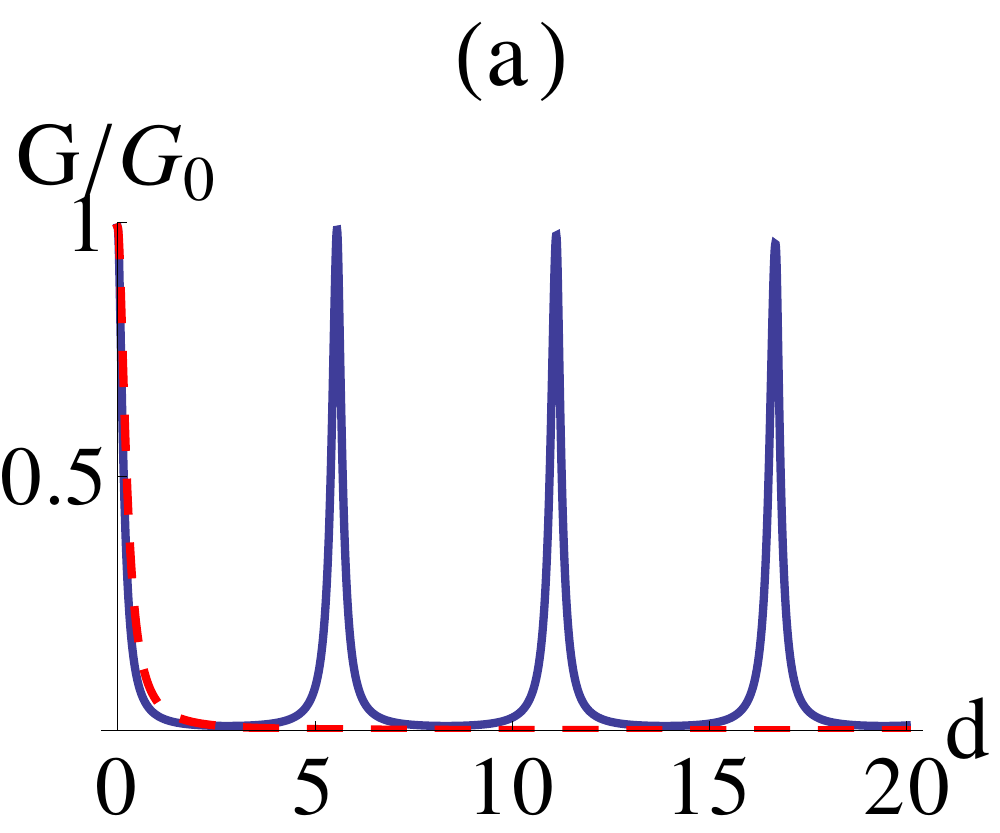}
\includegraphics[width=0.48 \linewidth]{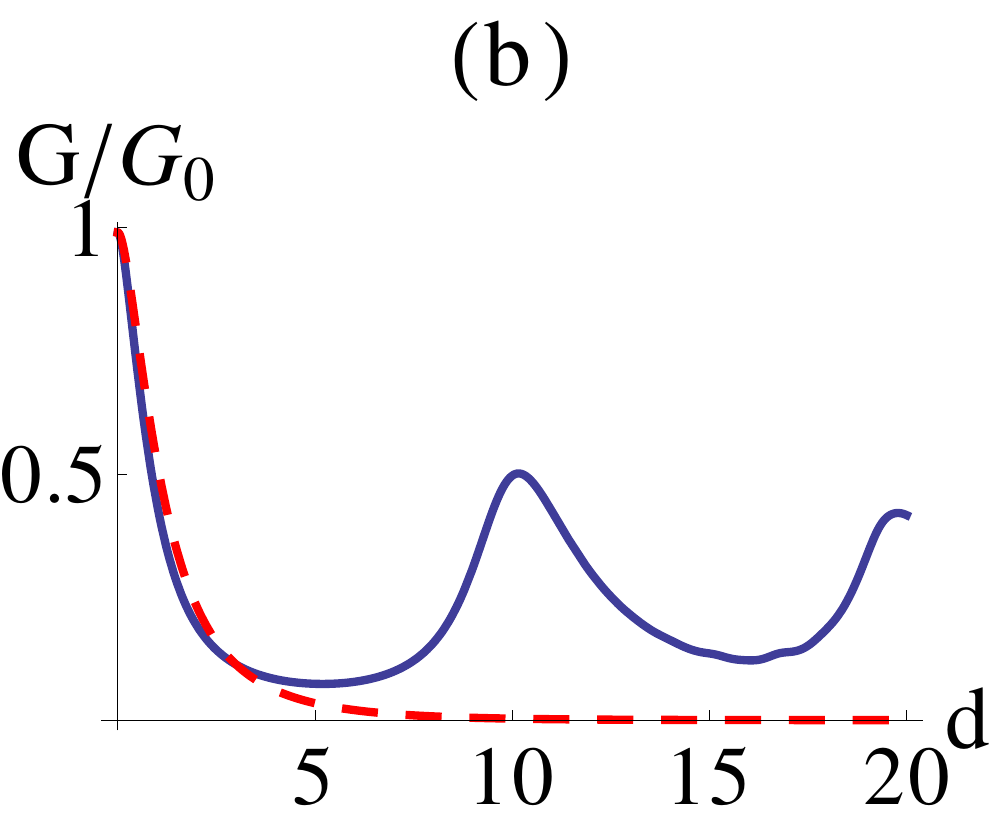}
\includegraphics[width=0.48 \linewidth]{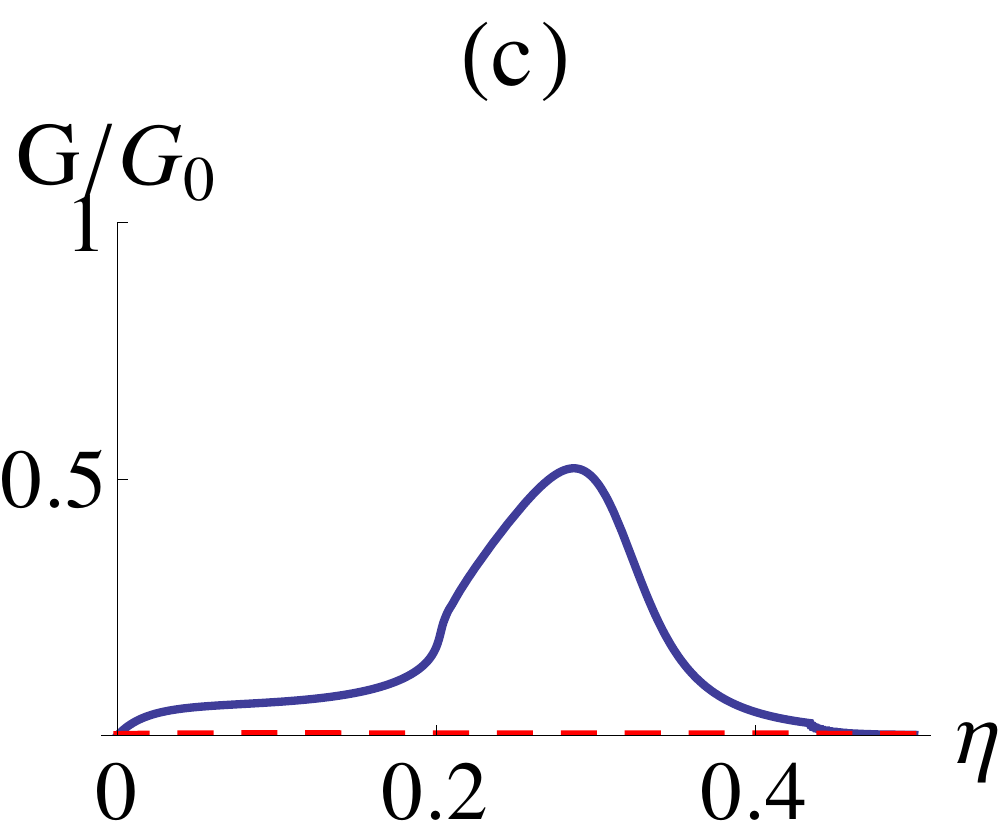}
\includegraphics[width=0.48 \linewidth]{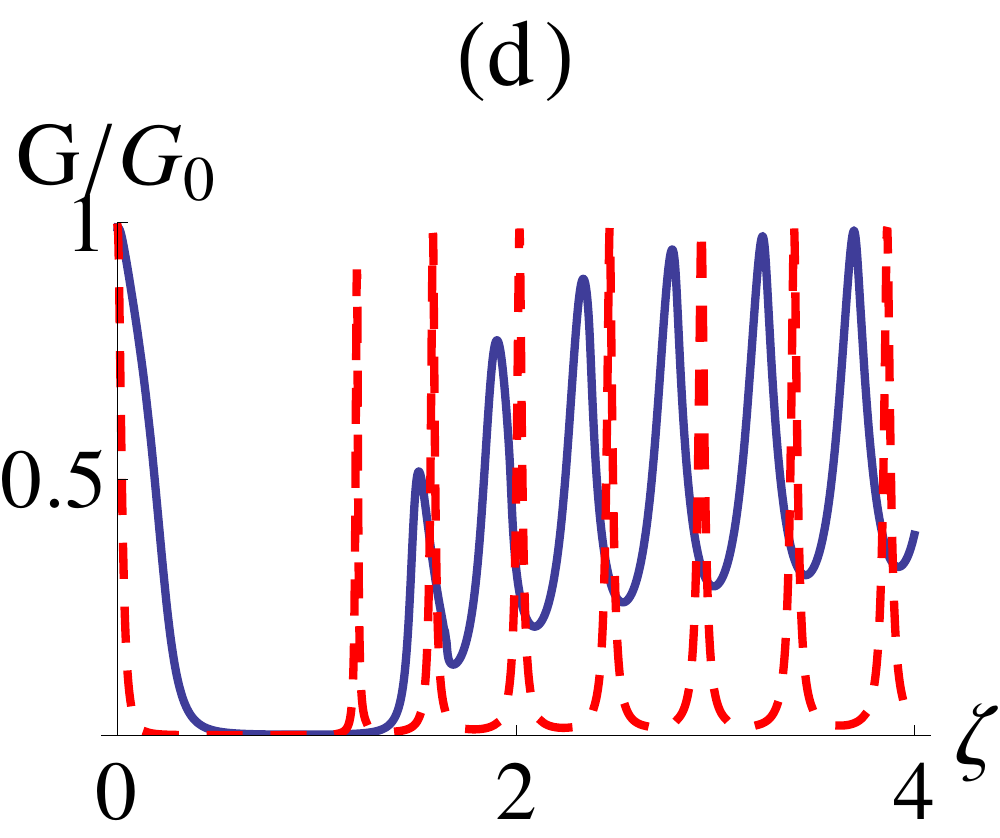}
\caption{ (a) Plot of the conductance $G/G_0$ as a function of $d$
for $\eta=0.01$, $\zeta=1.5 > \zeta_2$ (blue solid line) and
$\zeta=1< \zeta_2$ (red dashed line) displaying oscillatory and
decaying behavior respectively. (b) Similar plot for $\eta=0.3$ with
$\zeta=1.5 > \zeta_2$ (blue solid line) and $\zeta=1.2< \zeta_2$
(red dashed line). (c) Plot of $G/G_0$ as a function of the applied
voltage $\eta$ with $d=10$  for $\zeta=1.5$ (blue solid line) and
$1$ (red dashed line). (d) Plot of $G/G_0$ as a function of $\zeta$
with $d=10$ and several values of $\eta=0.3$ (blue solid line) and
$0.01$ (red dashed line). For $\eta=0.3(0.01)$, $\zeta_1=0.3(0.01)$
and $\zeta_2=1.3 (1.01)$. See text for details. \label{fig5}}
\end{center}
\end{figure}

Next, we plot the dimensionless conductance $G/G_0$ as a function of
the barrier width $d$ in Fig.\ \ref{fig5}(a). We find that (top left
panel) for small $\eta$, $G/G_0$, which mimics the behavior of
$T_1(k_y=0)$, shows oscillatory behavior for $\zeta > \zeta_2$; the
period of these oscillations are determined by $k'_x d= \pi$. For
smaller $\zeta < \zeta_2 =1 +\eta$, $G/G_0$ decays with barrier
strength; thus one can effectively tune the conductance of
phosphorene by tuning the strength of potential barriers along $x$,
provided that the applied voltage $\eta$ stays small. For larger
$\eta$, the behavior of $G$ deviates from that of $T_1$ since a
large number of $k_y$ modes contribute to the transmission; in this
regime $G/G_0$ (Eq.\ \ref{contcond1}) never reaches its maximal
value of unity. This feature is shown in Fig.\ \ref{fig5}(b);
however, we note that $G$ still remains an oscillatory or decaying
function of $d$ depending on the value of $\zeta$. The bottom panel
of Fig.\ \ref{fig5} displays the behavior of $G/G_0$ as a function
of the applied voltage $\eta$ (left panel) for fixed barrier
potential $\zeta$ and barrier width $d$ and as a function of $\zeta$
for fixed $\eta$ and $d$ (right panel).  For the former plot, in
Fig.\ \ref{fig5}(c), $G$ is initially an increasing function of
$\eta$ followed by a decaying behavior for $\zeta=1.5$ while for
$\zeta=1$, it stays close to zero for all $\eta$. This can be
understood from the fact that $\zeta_2=1+\eta$ is an increasing
function of $\eta$. If $\zeta> \zeta_2(\eta=0)=1$, as $\eta$ (and
thus $\zeta_2$) increases and crosses $\zeta$, the behavior of $G$
changes from oscillatory to decaying as a function of $d$. This is
reflected by an initial increase in $G$ as a function of $\eta$
followed by a decaying behavior for $\zeta<\zeta_2$.  However, if
$\zeta_1(\eta=0)=0 \le \zeta \le \zeta_2(\eta=0) =1$, $G$ is always
small and remains close to zero for any $\eta$; this happens for
$\zeta=1$ as shown in Fig.\ \ref{fig5}(c) (red dashed line). For the
latter plot, in Fig.\ \ref{fig5}(d), we find that $G$ decreases as a
function of $\zeta$ for a fixed $d$ for $\zeta_1 \le \zeta<\zeta_2$;
however, as $\zeta$ is increased to values larger than $\zeta_2$,
$G$ becomes an oscillatory function of $\zeta$. This behavior can be
easily understood from Eq.~\eqref{transcont} noting that $T$ and
therefore $G$  is an oscillatory or decaying function of $\zeta$
depending on whether $k'_x$ is real of imaginary. Since $k'_x$
becomes imaginary for $\zeta_1 \le \zeta \le \zeta_2$, $G/G_0$
becomes a decaying function of $\zeta$ within this range. For $\zeta
\gg \zeta_2$, $G/G_0$ oscillates with $\zeta$; thus we find that $G$
changes from being oscillatory to decaying to oscillatory function
of $\zeta$ with increasing $\zeta$.

\begin{figure}[t]
\begin{center}
\includegraphics[width=0.48 \linewidth]{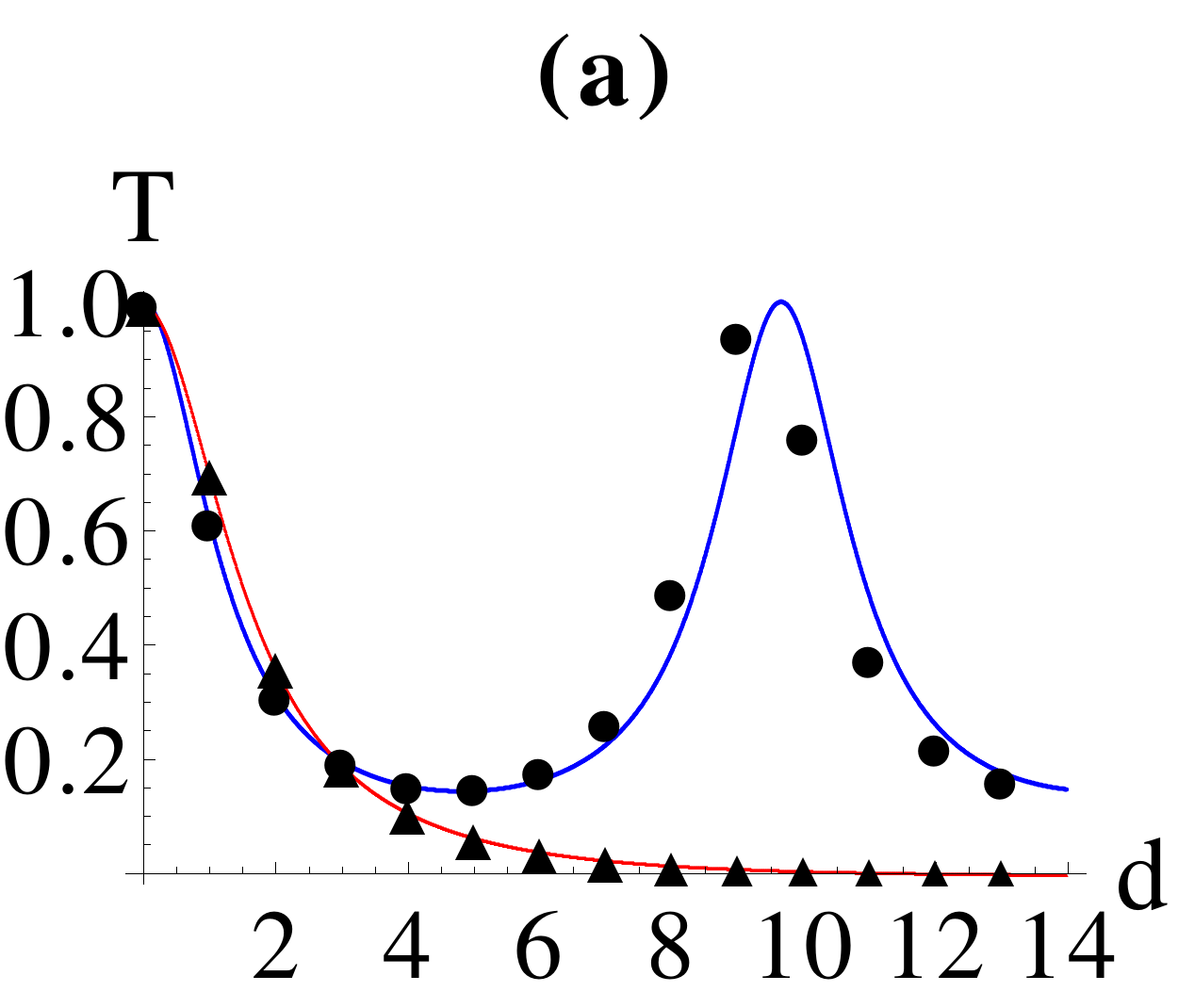}
\includegraphics[width=0.48 \linewidth]{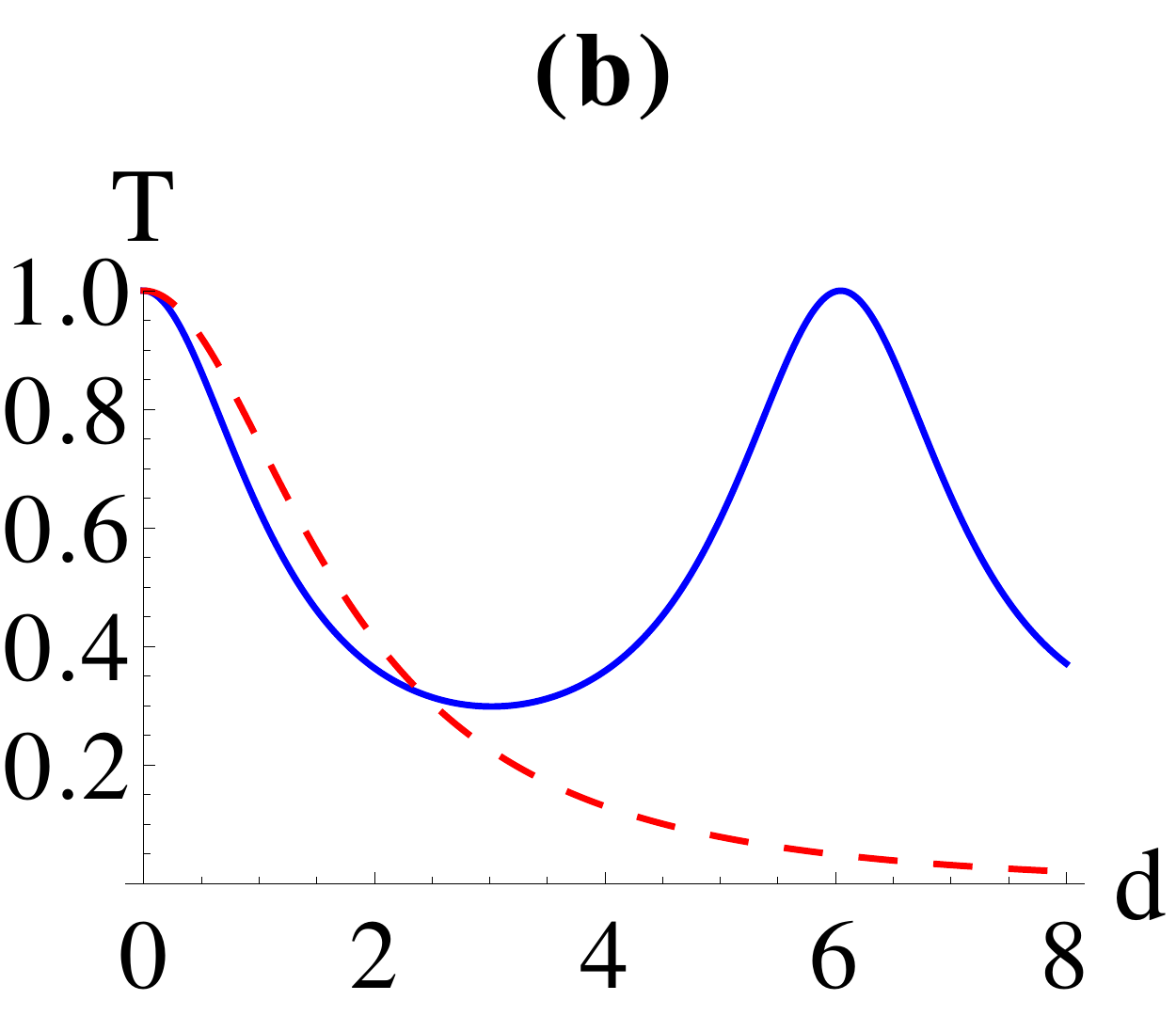}
\includegraphics[width=0.48 \linewidth]{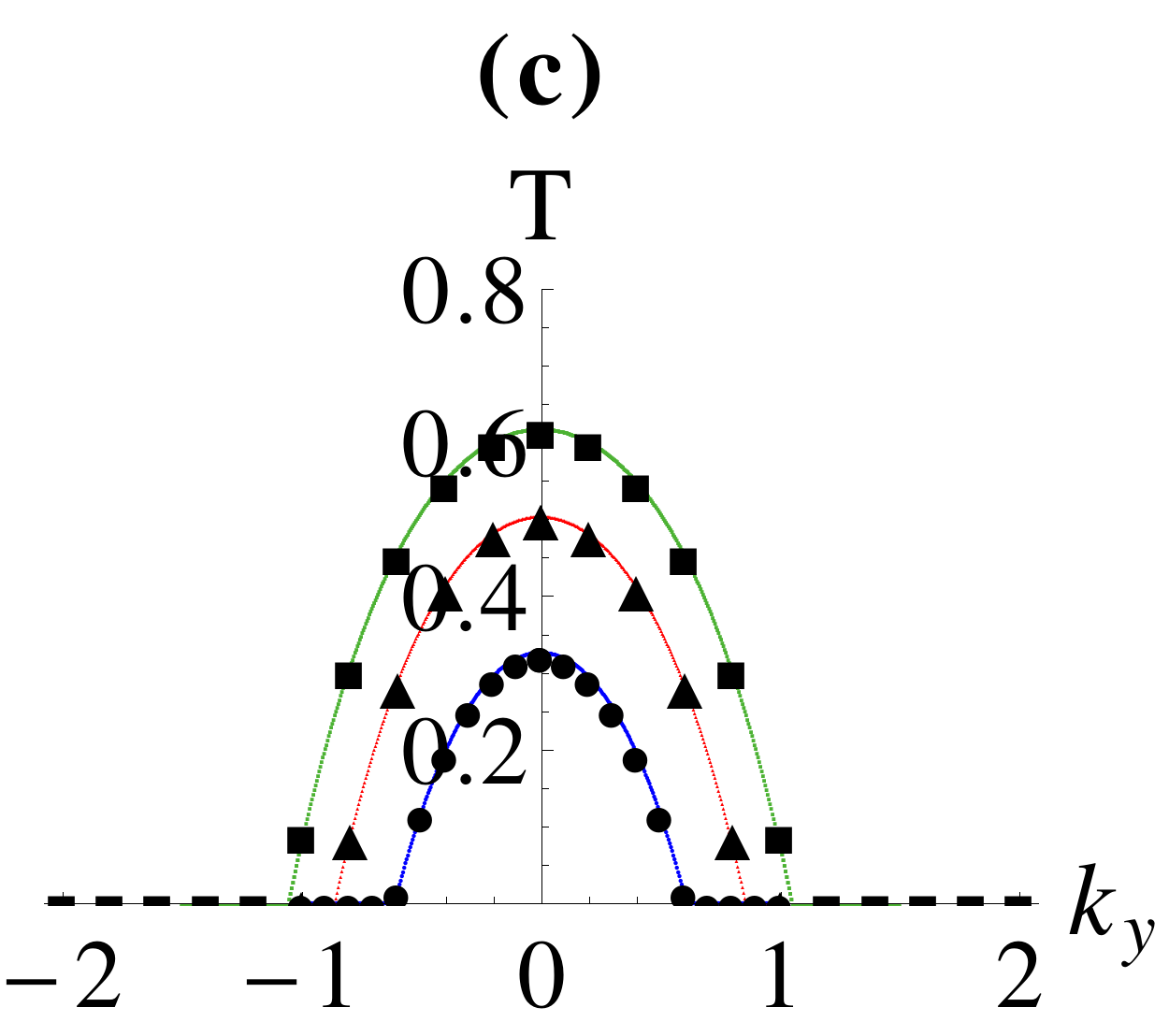}
\includegraphics[width=0.48 \linewidth]{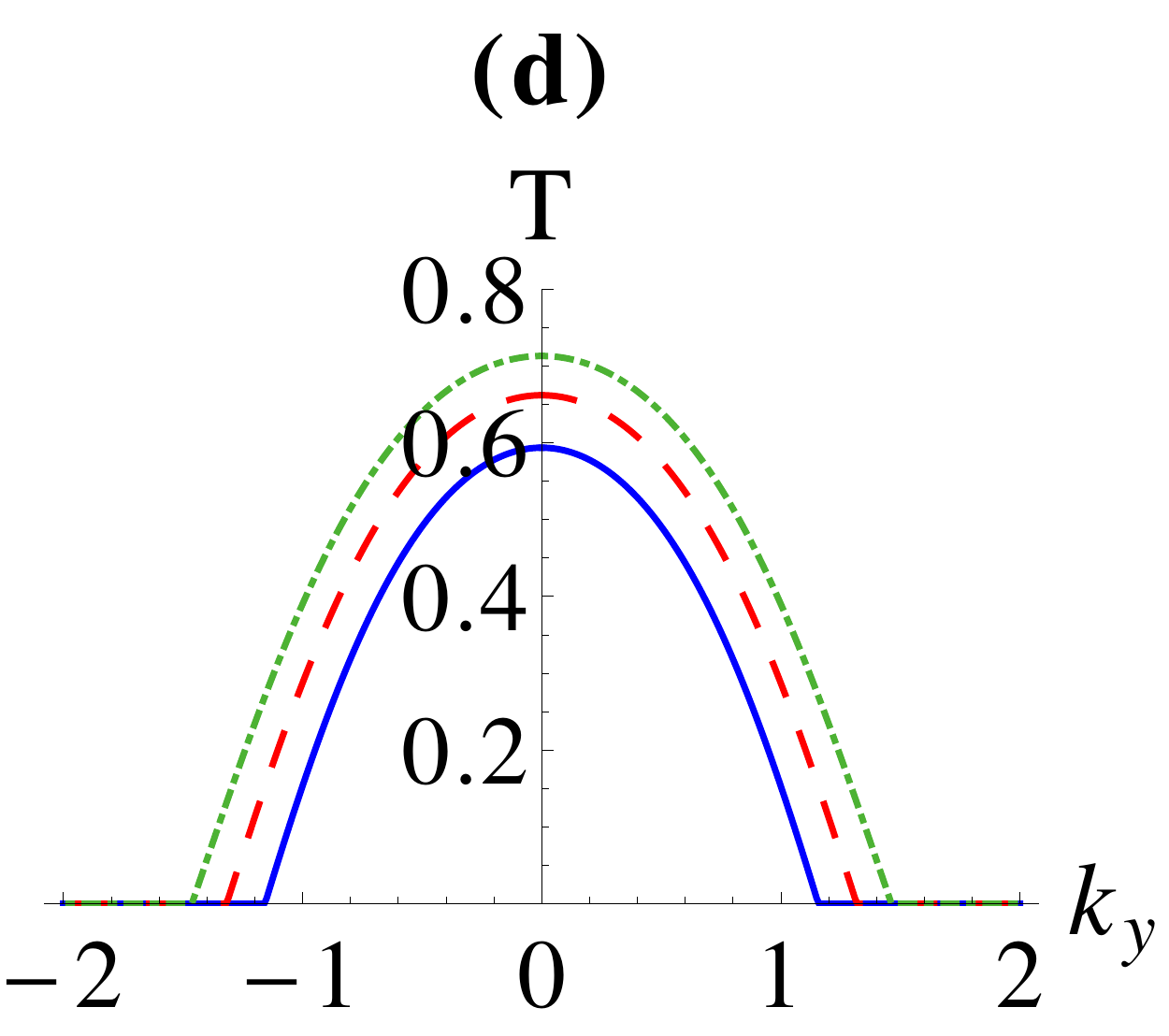}
\caption{ (a) Plot of the normal transmission $T\equiv T(k_y=0)$
computed from the lattice model as a function of $d$ for $\eta=0.3$
and $\zeta=1.5 > \zeta_2$ (blue solid line) and $\zeta=1.2< \zeta_2$
(red dashed line) displaying oscillatory and decaying behavior
respectively. The black circles ($\zeta=1.5$) and triangles
($\zeta=1.2$) are results obtained from the continuum calculation.
(b) Same as in (a) but for larger values of the applied voltage
$\eta=0.4$ where the continuum model is expected to be inaccurate.
The oscillatory behavior corresponds to $\zeta=1.8$ (blue solid
line) and the decaying behavior to $\zeta=1.3$ (red dashed line).
(c) Plot of the transmission $T(k_y)$ as a function of $k_y$
displaying collimated behavior for $d=1$, $\zeta=1.5$ and $\eta=0.1$
(blue line), $0.2$ (red line) and $0.3$ (green line). The
corresponding continuum results are plotted as black circle
($\eta=0.1$), triangle ($\eta=0.2$) and square ($\eta=0.3$). (d)
Plot of the transmission $T(k_y)$ as a function of $k_y$ displaying
collimated behavior for $d=1$, $\zeta=1.8$ and $\eta=0.35$ (blue
line) $0.45$ (red line) and $0.55$ (green line). See text for
details. \label{fig6} }
\end{center}
\end{figure}

Next, we compare the continuum results obtained with those obtained
from the lattice Hamiltonian (Eq.~\eqref{hamk1}) using
Eqs.~\eqref{latbc1} and \eqref{condlatxx}. A plot of the normal
transmission as a function of $d$ obtained from the lattice model is
compared to those obtained from the continuum model in Fig.\
\ref{fig6}(a); we find that the plots agree quite well to the
results of the continuum model for small $\eta$. For larger
$\eta>0.3$, where the continuum model is expected to fail, the
results of the lattice model, shown in Fig.\ \ref{fig6}(b), shows
the expected oscillatory (and decaying) behavior of $T(k_y=0)$ with
$d$ for $\zeta>(<)\zeta_2$. Thus we demonstrate that the change from
oscillatory to decaying behavior of transmission with $d$ found in
the continuum model holds at larger applied voltages where the
lattice model provides an accurate description of the transport. The
plot of $T$ as a function of the transverse momentum $k_y$, shown in
Fig.\ \ref{fig6}(c), reproduces the expected collimated nature of
the transmission; moreover, it agrees quite well with the results
obtained form the continuum model. This highlights the fact that as
long as $\eta$ is small, the results of the lattice and the
continuum models agree with each other; this provides the
justification of our continuum analysis which provides a better
analytic insight for the obtained results. In Fig.\ \ref{fig6}(d),
we show the nature of $k_y$ dependence of $T$ for larger applied
voltages, $\eta>0.3$, for which the continuum model is not expected
to be accurate.

\begin{figure}[t]
\begin{center}
\includegraphics[width=0.75\linewidth]{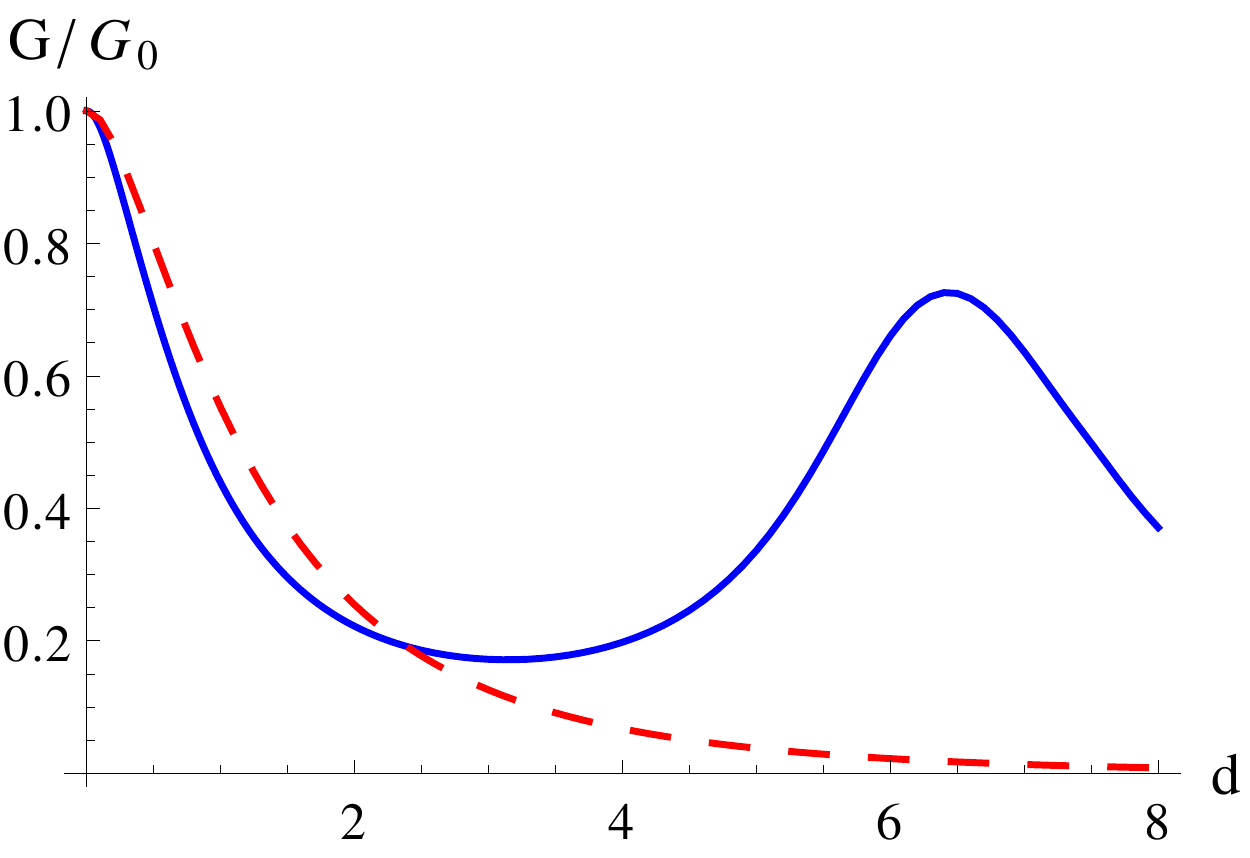}
\caption{ Plot of $G$ as a function $d$ for $\eta=\zeta_1=0.4$ and
$\zeta=1.8 >\zeta_2=1.4$ (blue solid line) and $\zeta=1.3 <\zeta_2$
(red dashed line). See text for details \label{fig7n}}
\end{center}
\end{figure}

Finally, we note that the oscillatory/decaying behavior of $T$ as a
function of $d$ can also be found in $G$, as calculated from the
lattice model, for $\eta > 0.3$. This is shown in Fig.\ \ref{fig7n};
we find that $G$ displays oscillatory (decaying) behavior as a
function of $d$ for $\zeta>\zeta_2=1.4$ ($0.4=\zeta_1\le \zeta \le
\zeta_2$). This demonstrates that the switch from oscillatory to
decaying behavior, evident from the continuum model valid for small
applied voltages, also persists for a larger ranges of applied voltages
$\eta$ where the continuum model is not expected to be accurate;
this property is therefore more general, and not a consequence of
the  continuum approximation used earlier in this section. This
property stems from the longitudinal wavevector in the barrier
switching from real to imaginary which occurs for both the lattice
and the continuum model.

\begin{figure}[t]
\begin{center}
\includegraphics[width= 0.75\linewidth]{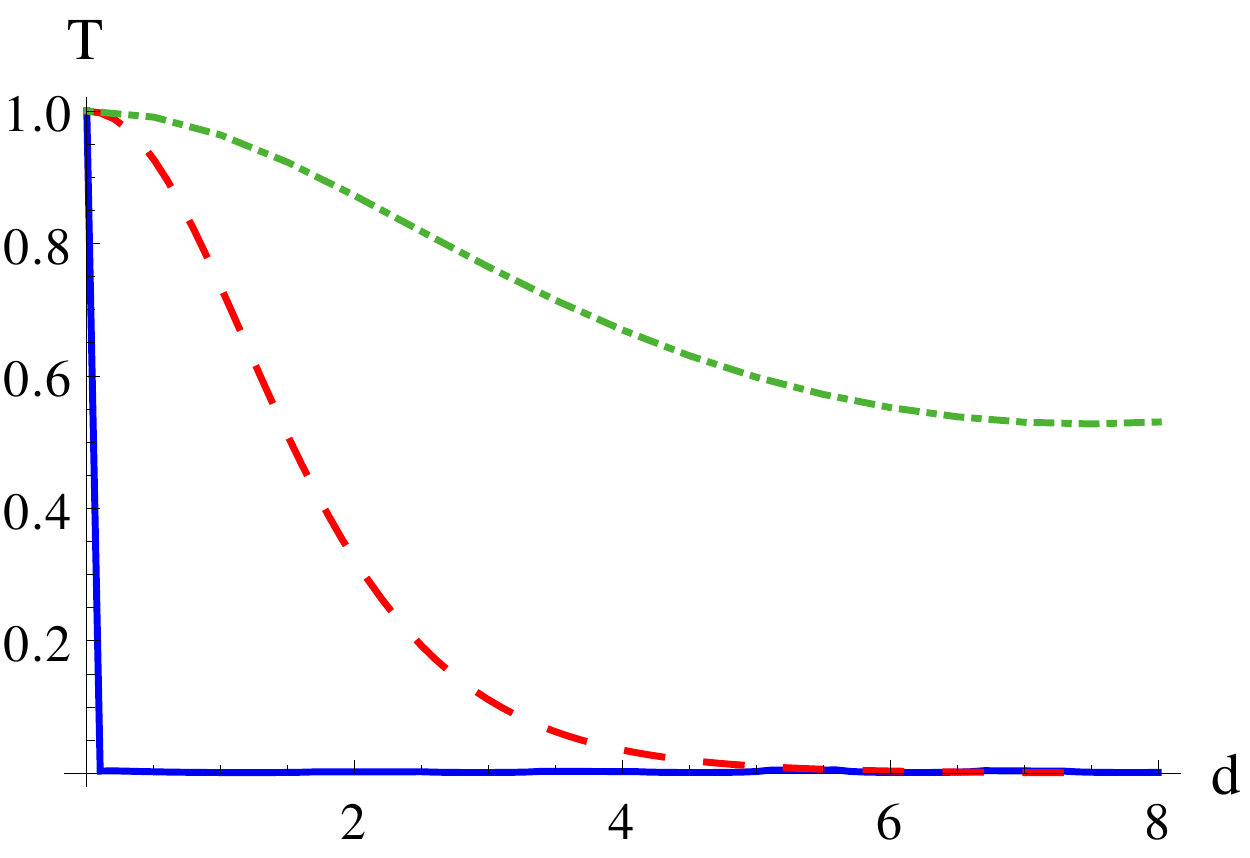}
\caption{ Plot of $T\equiv T(k_x=\pi/24)$ as a function $d$ for
barrier along $y$ with $\eta=0.1$ and for $\zeta=1.8$ (blue solid
line) $0.15$ (red dashed line) and $0.05$ (green dash-dotted line).
See text for details.  \label{fig8n}}
\end{center}
\end{figure}

Having studied the transport across a barrier oriented along the $x$
(armchair) direction, we now compare the transmission for a single
barrier along $y$ (zigzag) direction with height $U_0$ and width
$d$. The details of computing the transmission is analogous to the
procedure outlined in Sec.\ \ref{form1} for the lattice model with
the difference that $k_x$ now plays the role of transverse momentum.
The result of this analysis is shown in Fig.\ \ref{fig8n}. We find
that in contrast to the transmission along $x$, $T$ displays
decaying behavior as a function of $d$ for all $\zeta$ as expected
for transport mediated by Schr\"odinger quasiparticles with a
parabolic dispersion. This difference originates from the
anisotropic band structure of phosphorene and highlights the
importance of the orientation of the barrier which controls the
qualitative nature of transport in this material.

\section{Multiple NBN junctions}
\label{sec3}

In this section, we shall extend the formalism developed in Sec.\
\ref{form1} to treat the case of multiple potential barriers along
$x$. The proposed experimental setup is sketched in Fig.\
\ref{fig1}(d). We shall consider the case where all the barriers have a
height $U_0$ and width $d$; the distance between two successive
barriers is denoted by $b$. In what follows, we are going to use the
continuum approximation discussed in Sec.\ \ref{sec1} so that the
system Hamiltonian is given by Eq.~\eqref{tbh1}.

To calculate the transmission across multiple barriers, we use the
transfer matrix formalism which has been developed for gapless Dirac
electrons on the surface of a topological insulator in
Ref.~[\onlinecite{smondal1}]. The first step in developing this
formalism constitutes obtaining an expression for the wavefunction
in the $l=(2n+1)^{\rm th}$ region ({\it i.e.} region between $l-1$
and $l+1$ barriers), where $n$ is an integer.  Recall that the
barrier regions are labeled by even integers in Fig.~\ref{fig1}(d). The
quasiparticle wave function in different regions are given by
$\Psi_{l}=\psi_{l} e^{i k_y
y}/\sqrt{2}$, and $\psi_l$ 
is given by \cite{smondal1}
\begin{eqnarray}
\psi_{l} &=& \mathcal{G} M(x) A_l, \quad l = (2n+1),
\nonumber\\
&=& \mathcal{G}' M'(x) A_l, \quad l=2n. \label{mbwav1}
\end{eqnarray}
Here the coefficients are given by $A_i=(a_i\ \
b_i)^T$ and the matrices $G$ and $M$ are given by
\begin{equation}
  \mathcal{G}=\begin{pmatrix}
    1&1\\
    \lambda_0 e^{i\phi_{\kk}}&\lambda_0 e^{-i\phi_{\kk}}
  \end{pmatrix}
  ,\ \
   M(x)=\begin{pmatrix}
   e^{i k_x x}&0\\
    0&e^{-i k_x x}
  \end{pmatrix}, \label{pmat1}
 \end{equation}
and $\mathcal{G}'(M'(x))$ is obtained from $\mathcal{G}(M(x))$ by
replacing $k_x$ with $k'_x$ along with $\lambda_0 \to \lambda_0'$.

Next, we employ the boundary conditions at the junctions by imposing
continuity of the wavefunctions across the barrier. This leads to,
after some algebra, the expressions for the reflection ($r$) and
transmission ($t$) coefficients across the barriers. The reflection and the transmission coefficients
can be expressed in terms of the elements of the transfer matrix $N$ as
\begin{eqnarray}
t&=& N_{22}^{-1}, \quad r= -N_{21}/N_{22} \label{xcoeff}
\end{eqnarray}
where $N$ (for a system with n barriers) is defined via the relation $A_{2n+1}=NA_{1}$. It can be evaluated to be
\begin{eqnarray}
N &=& N_{n}.N_{n-1}....N_{2}.N_{1}~,~~~~{\rm where}~ \no\\
N_i &=&\left[\mathcal{G}_1.M_1(x_{i2})\right]^{-1}.
\mathcal{G}_2.M_2(x_{i2}). \nonumber\\
&& \left[\mathcal{G}_2.M_1(x_{i1})\right]^{-1}.
\mathcal{G}_1.M_1(x_{i1}), \label{nweq1}
\end{eqnarray}
where $\ x_{ij}=(i-1)(d+b)+(j-1)d$ and the elements of the matrix
$N_i$ are explicitly given by
\begin{eqnarray}
N_i&=&\left(\begin{array}{c c}
    w & z^\ast_i\\
    z_i & w^\ast
   \end{array}\right)
,~~~~{\rm where}~\no\\
w&=&e^{-ik_xd}\frac{|B_1|^2 e^{-ik_x'd}-|B_2|^2 e^{ik_x'd}}
{{\rm Det}[\mathcal{G}_1]{\rm Det}[\mathcal{G}_2]},\no\\
z_i&=&\frac{-2i B_1B_2\sin (k_x'd) e^{ik_x(x_{i1}+x_{i2})}} {{\rm
Det}[\mathcal{G}_1]{\rm Det}[\mathcal{G}_2]},  \label{nweq2}
\end{eqnarray}
and $k_x$ and $k_x'$ are defined in Eq.~\eqref{kxcont} and
Eq.~\eqref{wavecont2}, respectively. The transmission coefficient
$T$ and the conductance $G$ across the barrier can be obtained as
$T= |t|^2$ and $G = G_0 \int dk_y T(k_y)/(2\pi)$.

\begin{figure}[t]
\begin{center}
\includegraphics[width= 0.75\linewidth]{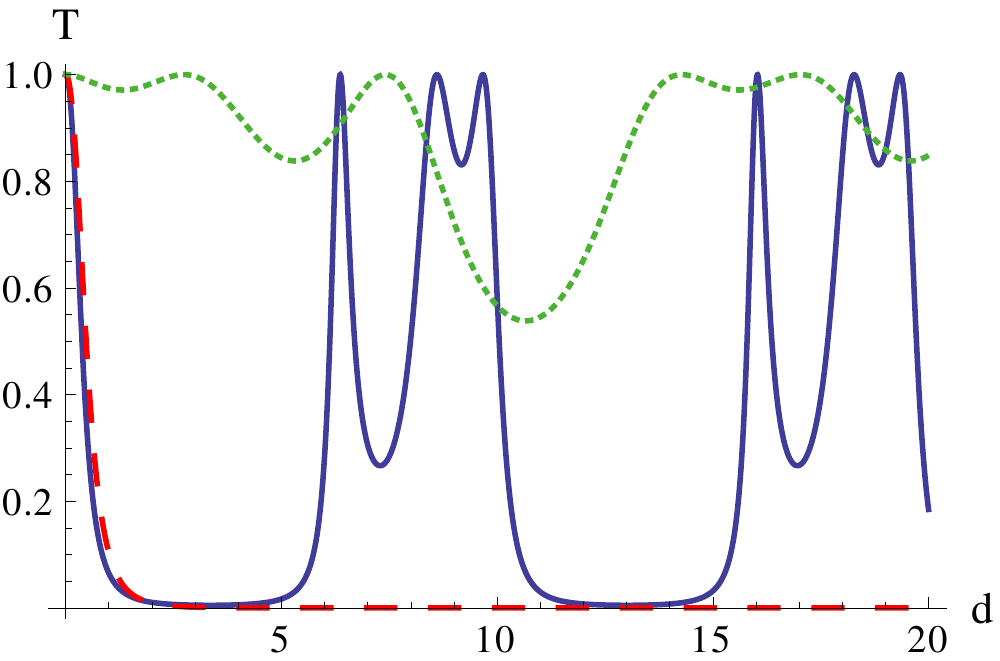}
\caption{ Plot of the normal transmission $T(k_y=0)$ as a function
of $d$ for $\eta=0.3$, $n=3$, $b=1$ and $\zeta=1.5>\zeta_2=1.3$
(blue solid line), $0.3=\zeta_1 \le \zeta=1.2 \le \zeta_2$ (red
dashed line) and $\zeta=0.3 \le \zeta_1$ (green dotted line). The
number of barriers is set to $n=3$. \label{fig7}}
\end{center}
\end{figure}

\begin{figure}[t]
\begin{center}
\includegraphics[width=0.48 \linewidth]{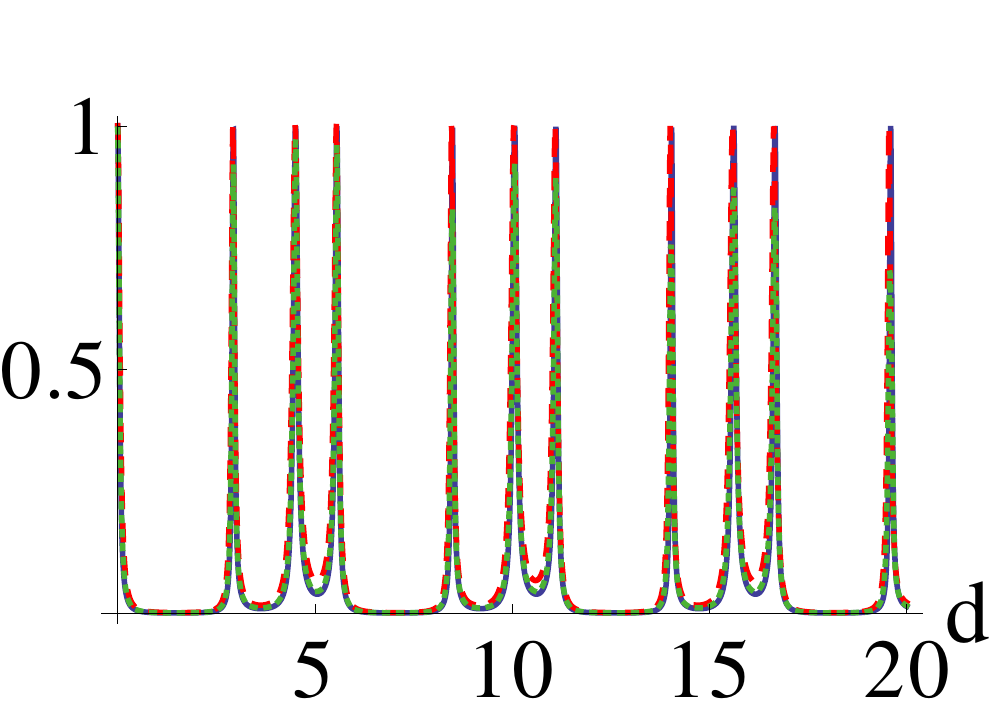}
\includegraphics[width=0.48\linewidth]{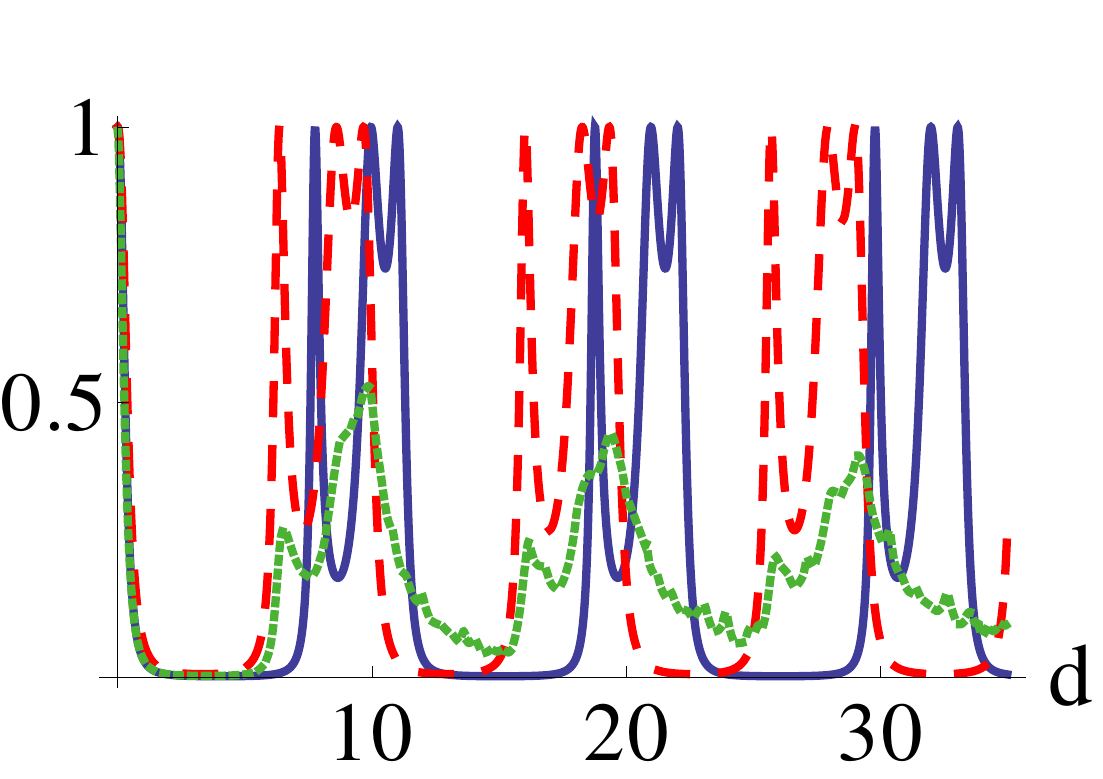}
\caption{ Plot of the conductance $G/G_0$ and transmission $T(k_y)$
for representative $k_y$ values as a function of $d$ for $n=3$ and
$\eta=0.01$ (left panel) and $0.3$ (right panel). In the left panel,
$T(k_y=0)$ (blue solid line), $T(k_y=\pi/25)$ (red dashed line) and
$G/G_0$ (green dotted line). In the right panel, $T(k_y=0)$ (blue
solid line), $T(k_y=\pi/6)$ (red dashed line) and $G/G_0$ (green
dotted line). For both plots, all other parameters are same as in
Fig.\ \ref{fig6}. We note that for small $\eta$, $G/G_0$ mimics
$T(k_y=0)$; however, the behavior of these quantities are quite
different for larger $\eta$, as obtained in the case of a single
barrier. \label{fig8}}
\end{center}
\end{figure}

\begin{figure}[t]
\begin{center}
\includegraphics[width= 0.75\linewidth]{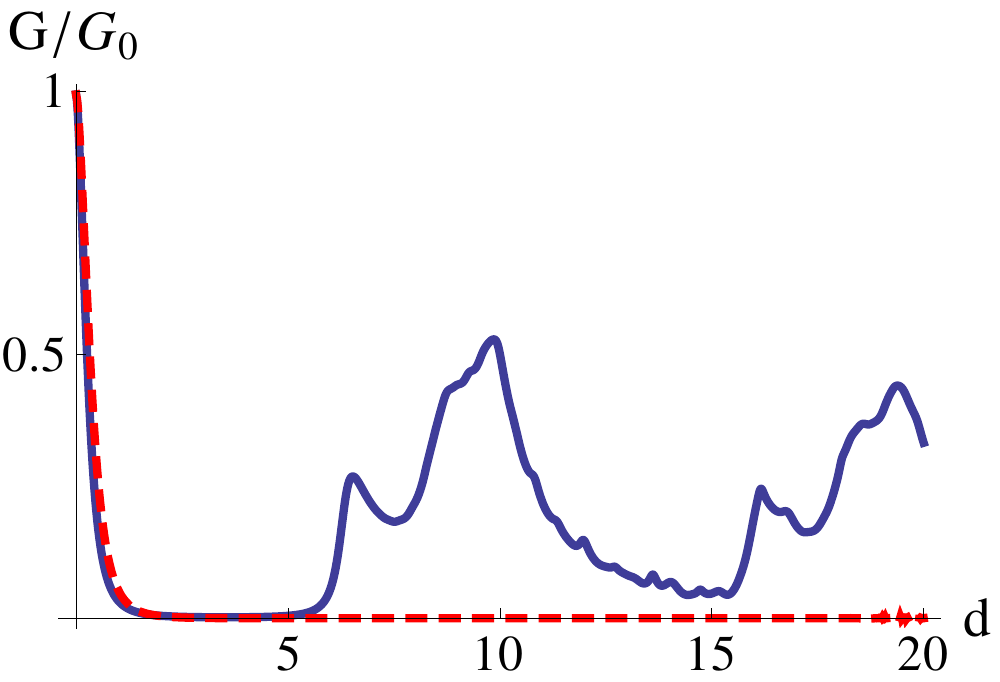}
\caption{ Plot of the conductance $G/G_0$ as a function of $d$ with
$\eta=0.3$ for $\zeta=1.5 >\zeta_2=1.3$ (blue solid line) and
$\zeta=1.2<\zeta_2$ (red dashed line). All other parameters are same
as in Fig.\ \ref{fig6}.\label{fig9}}
\end{center}
\end{figure}

To explore the impact of multiple barriers, we first plot the normal
transmission $T(k_y=0)$ for low applied voltage $\eta$ as a function
of $d$ for $n=3$ barriers in Fig.\ \ref{fig7}. We find that similar
to the behavior of $T$ for the single barrier case, for multiple
barriers, $T$ is either a decaying or an oscillatory function of $d$
depending on the value of $\zeta$. The only quantitative difference
that we find are as follows. First, the structure of the
oscillations for $\zeta > \zeta_2$ or $\zeta<\zeta_1$ becomes more
complicated due to the presence of multiple barriers and second the
decay of $T$, for $\zeta_1 \le \zeta \le \zeta_2$ becomes much
sharper. Next, we compare the behavior of conductance $G$ with that
of $T$ for several representative values of $\zeta$ in
Fig.~\ref{fig8}. We note that as in the case of single barrier, $G$
mimics $T$ for small $\eta$ (left panel of Fig.\ \ref{fig8}); for
larger values of $\eta$, $G$ and $T$ shows different behavior. This
indicates collimated transport for small $\eta$ across multiple
barriers, which is also expected from our results in Sec.\
\ref{result1}. Finally, we plot the conductance $G$ as a function of
$d$ for $\eta=0.3$ and two representative values of $\zeta$ in Fig.\
\ref{fig9}. We find that for $\zeta \le \zeta_2$, $G$ shows a sharp
decay as a function of $d$ while it shows oscillatory behavior for
$\zeta > \zeta_2$.

The nature of the oscillations of transmission $T$ or conductance
$G$ can be understood analytically by considering the simplest case
of multiple barriers corresponding to $n=2$. In this case, some
straightforward algebra, starting from Eqs.\ \eqref{pmat1},
\eqref{nweq1}, and \eqref{nweq2}, yields an analytic expression of $T_2
\equiv T^{n=2}$ as
\begin{eqnarray}
T_2&=& \Big [T_1^{-2}+(1-T_1^{-1})^2-2(1-T_1^{-1}) \nonumber\\
&& \times {\rm Re}[w^2 e^{2ik_x (d+b)}]\Big]^{-1},  \label{n2trans}
\end{eqnarray}
where $T_1$ is the transmission through a single barrier. From
Eq.~\eqref{n2trans}, it is easy to see that $T_2=1$ for $T_1=1$
({\it i.e} for $k'_x d= m \pi$, where $m$ is an integer) and also
for  $T_1=-1/{\rm Re}[w^2 e^{2ik_x (b+d)}]$ leading to two distinct
peaks of the transmission. The number of such peaks increase with
$n$; thus the oscillation pattern becomes more complicated with
increasing $n$. We also note that the behavior of $T_2$ as a
function of $d$ is controlled by $T_1$; hence the oscillatory or
decaying nature of the transmission and conductance, across multiple
barriers, is also controlled by whether $k'_x$ becomes real of
imaginary for a given $\eta$, $\zeta$ and $k_y$.


\section{Discussion}
\label{sec4}

In this work, we have studied transport through a single as well as
multiple barriers in phosphorene along the armchair edge
($x$-direction) and have shown that such transport have features
which are qualitatively distinct from their counterparts for both
conventional Schr\"odinger materials and gapless Dirac systems such
as graphene and topological insulator surfaces. We have also shown
that due to the anisotropic band structure in phosphorene, such
unconventional properties are expected only for barriers along $x$;
for barriers along $y$, the transport displays standard
Schr\"odinger behavior.

The key unconventional features that our study unravel are as
follows. First, we show that the band structure of phosphorene
allows near normal transmission to dominate the conductance as the
applied voltage is tuned to near the bottom of the conduction band.
Consequently one obtains progressively collimated transport as $V$
approaches the band bottom. Second, in the limit where the applied
voltage is close to the conduction band bottom, the behavior of $G$
is identical to that of $T(k_y=0)$. It is well-known that the
oscillation of normal transmission with the barrier width/height
indicates signature of Klein paradox in graphene. However in
realistic experimental situation, one always measures $G$; thus the
analogous behavior of $T(k_y=0)$ and $G$ makes phosphorene an ideal
candidate for observing Klein paradox in Dirac materials via
conductance measurement. Both of these features stem from the
relatively flat band of phosphorene along the zigzag edge ($y$ or
$\Gamma-Y$) as opposed to the armchair edge ($x$ or $\Gamma-X$) --
see Fig.~{\ref{fig1}(b)}. Third, we find that in contrast to gapless
Dirac systems, both the normal transmission $T$ and the conductance
$G$ in single layer phosphorene displays a switch to monotonically
decaying function of the barrier width $d$ (from being an
oscillatory one) for a range of the applied voltage $\eta=\zeta_1
\le \zeta \le \zeta_2=\eta+1$ (where $\eta=eV/2m$ and $\zeta=
U_0/2m$), when the wavevector in the barrier region becomes
imaginary. Such a behavior also manifests itself in the dependence
of $G$ on $\eta$ and $\zeta$ and can thus be easily observed in
standard experiments. The origin of this behavior stems from both
the gapped Dirac structure and specific band structure of
phosphorene; it has not been observed in other known Dirac
materials. We have also shown that such a switch does not occur for
barriers along $y$; in that case, the behavior of $G$ behaves as
that due to conventional Schr\"odinger electrons (always decaying
with increasing barrier width). This dichotomy arises due to
unconventional band structure of phosphorene. Finally, our analysis
of the multiple graphene junctions shows that the unconventional
behavior of $G$ mentioned above, persists for multiple barriers
along $x$;  the oscillations of $G$ or $T(k_y=0)$ in the regime
$\zeta_1\ge \zeta \ge \zeta_2$ becomes more complex and develops
additional peaks. The reason for this has been analytically
explained for $N=2$.

Experimental verification of our work would require constructing
potential barrier/barriers of height $U_0$ and width $d$ atop single
layer phosphorene whose chemical potential is set to the bottom of
the conduction band by application of external bias. This has
already been done for graphene \cite{mandar1} and analogous
procedures may be used here. We propose conductance measurement of
such a system as a function of the applied bias voltage $V$; our
theory predicts that $G$ would mimic the behavior shown in Fig.\
\ref{fig5}(c) and (d) as a function of the applied bias voltage $V$
and barrier height $U_0$.  Further the oscillatory behavior of $G$
as a function of $\chi_0$ (which can be varied by varying $U_0$ or
$d$) for small $\eta$ and with $\zeta>\zeta_2=1+\eta$ as shown in
Figs.\ \ref{fig5}(a) and \ref{fig6}(a) would display signature of
Klein paradox for gapped Dirac systems. In this context we note that
unlike graphene, the band structure of multi-layer phosphorene
remains qualitatively similar to that of single layer, though with
different parameters. Thus the transport properties of a NBN
junction in multi-layer phosphorene is also expected to be
similar to the one discussed in this work \cite{Barun}.


In conclusion, we have studied the transport properties of single
layer phosphorene in the presence of single and multiple barriers
along $x$; our results show unconventional properties of such
transport which are different from those of both conventional
electron gas and gapless Dirac material observed in graphene. We
also find that the band structure of phosphorene allows for tuning
into a regime where the transport is naturally collimated and the
conductance is dominated by contribution of normal transmission
across the barrier; this in turn establishes phosphorene as a likely
experimental platform for observing Klein paradox. We have discussed
experiments which are likely to unravel these unconventional
features.

\vspace{0.1cm}

\section*{Acknowledgement}

AA acknowledges funding support from the DST INSPIRE Faculty Award.

\end{document}